\documentclass[fleqn,usenatbib]{mnras}

\usepackage[T1]{fontenc}
\usepackage{ae,aecompl}
\usepackage{newtxtext,newtxmath}

\DeclareRobustCommand{\VAN}[3]{#2}
\let\VANthebibliography\thebibliography
\def\thebibliography{\DeclareRobustCommand{\VAN}[3]{##3}\VANthebibliography}

\usepackage{graphicx}	
\usepackage{amsmath}	
\usepackage{amsfonts}	
\usepackage[dvipsnames]{xcolor} 
\usepackage[labelfont=bf]{subcaption} 
\usepackage[flushleft]{threeparttable}
\usepackage{hyperref}
\usepackage{enumerate}
\usepackage{siunitx}
\usepackage{comment}
\usepackage{bm}

\newcommand{\twoi}{\mathrm{II}}
\newcommand{\threei}{\mathrm{III}}

\newcommand{\simcode}[0]{{\textsc{21cmSPACE}}}
\newcommand{\polychord}[0]{\textsc{PolyChord}}
\newcommand{\anesthetic}[0]{\textsc{anesthetic}}
\newcommand{\globalemu}[0]{\textsc{globalemu}}

\defcitealias{2021AJ....162...47B}{B21}
\defcitealias{2024MNRAS.533.3222D}{D24}

\title[Multi-wavelength constraints on 21-cm signal]{Narrowing the discovery space of the cosmological 21-cm signal using multi-wavelength constraints
}

\author[Dhandha et al.]
{Jiten Dhandha,$^{1,2}$\thanks{E-mail: \href{mailto:jvd29@cam.ac.uk}{jvd29@cam.ac.uk}}
Anastasia Fialkov,$^{1,2}$
Thomas Gessey-Jones,$^{2,3,6}$ 
Harry T. J. Bevins,$^{2,3}$ 
Sandro Tacchella,$^{2,3}$\newauthor
Simon Pochinda,$^{2,3}$
Eloy de Lera Acedo,$^{2,3}$
Saurabh Singh,$^{4}$
and Rennan Barkana,$^{5}$
\\
$^{1}$Institute of Astronomy, University of Cambridge, Madingley Road, Cambridge CB30HA, UK\\
$^{2}$Kavli Institute for Cosmology, Madingley Road, Cambridge CB30HA, UK\\
$^{3}$Astrophysics Group, Cavendish Laboratory, J. J. Thomson Avenue, Cambridge CB30HE, UK\\
$^{4}$Raman Research Institute, Bangalore, Karnataka 560080, India \\
$^{5}$School of Physics and Astronomy, Tel-Aviv University, Tel-Aviv 69978, Israel \\ 
$^{6}$PhysicsX, Victoria House, 1 Leonard Circus, London EC2A 4DQ, UK
}

\date{Accepted XXX. Received YYY; in original form ZZZ}

\pubyear{2025}

\begin{document}
\label{firstpage}
\pagerange{\pageref{firstpage}--\pageref{lastpage}}
\maketitle

\begin{abstract}
The cosmic 21-cm signal is a promising probe of the early Universe, owing to its sensitivity to the thermal state of the neutral intergalactic medium (IGM) and properties of the first luminous sources. Here, we constrain the 21-cm signal and infer IGM properties using the Population II galaxy parameters derived in a previous study through multi-wavelength synergies. This includes high-redshift UV luminosity functions (UVLFs) from Hubble Space Telescope (HST) and James Webb Space Telescope (JWST), cosmic X-ray and radio backgrounds (CXB and CRB), the SARAS~3 global 21-cm signal non-detection, and HERA 21-cm power spectrum upper limits. From CXB and HERA data, we infer the IGM kinetic temperature to be $T_\text{K}(z=15)\lesssim\SI{7.7}{K}$, ${\SI{2.5}{K} \lesssim T_\text{K}(z=10) \lesssim \SI{66}{K}}$, and ${\SI{20}{K} \lesssim T_\text{K}(z=6) \lesssim \SI{2078}{K}}$ at 95\% credible interval (CI). Similarly, CRB and HERA data limit the radio emission efficiency of galaxies, giving $T_\text{rad}(z=15) \lesssim \SI{47}{K}$, $T_\text{rad}(z=10)\lesssim\SI{51}{K}$, and $T_\text{rad}(z=6)\lesssim\SI{101}{K}$. These constraints, strengthened by UVLFs from HST and JWST, enable the first \textit{lower bound} on the cosmic 21-cm signal. We infer an absorption trough of depth ${\SI{-201}{mK}\lesssim T_\text{21,min} \lesssim \SI{-68}{mK}}$ at $z_\text{min}\approx10-16$, and a power spectrum of $\SI{8.7}{mK^2} \lesssim \Delta_{21}^2(z=15) \lesssim \SI{197}{mK^2}$ at $k=0.35\,h\text{Mpc}^{-1}$. Our results highlight the power of multi-wavelength synergies in constraining the early Universe. While promising for upcoming 21-cm experiments, the results depend on our assumption of a redshift-independent X-ray and radio efficiency of galaxies, and the exclusion of a flexible model for Population III stars.
\end{abstract}



\section{Introduction}

The 21-cm signal, tracing the evolution of neutral hydrogen across cosmic time, has long been recognized as a promising probe of the early Universe since its earliest theoretical predictions \citep[e.g.][]{1977SvAL....3..155V,1979MNRAS.188..791H,1990MNRAS.247..510S,1997ApJ...475..429M,2000ApJ...528..597T}. The evolution of the global (sky-averaged) 21-cm signal is predicted to pass through distinct phases: an initial absorption trough due to collisional coupling in a cooling intergalactic medium (IGM) during the Dark Ages \citep[e.g.][]{2000ApJ...528..597T,2004PhRvL..92u1301L}; a deeper absorption feature driven by Lyman-alpha coupling \citep[the Wouthuysen-Field effect;][]{1952AJ.....57R..31W,1958PIRE...46..240F} during the Cosmic Dawn, sourced by early metal-free Population~III stars \citep[e.g.][]{2015MNRAS.448..654Y,2022MNRAS.516..841G} or early Population~II galaxies \citep[e.g.][]{2004ApJ...602....1C,2005ApJ...626....1B}; weakening of the signal due to heating of the neutral intergalactic medium (IGM) by X-ray sources \citep[e.g.][]{2007MNRAS.376.1680P,2008ApJ...684...18C}; and eventual disappearance of neutral hydrogen driving the signal towards zero during the Epoch of Reionization \citep[EoR; e.g. see reviews by][]{2006PhR...433..181F,2012RPPh...75h6901P,2019cosm.book.....M}. However, the strength and timing of these phases are highly sensitive to the physical properties of the early Universe, and uncertainties in key parameters—such as the star formation efficiency (SFE), X-ray luminosity of high-redshift galaxies, and possible excess radio background, translate into a wide range of plausible 21-cm signal models.

The field has seen a rapid growth in the past decade, both on the theoretical and observational fronts. The reported EDGES detection of a deep absorption signal at $\nu\sim\SI{78}{MHz}$ \citep{2018Natur.555...67B} spurred a large number of theoretical studies focused on understanding the implications of its unusual depth and spectral shape. Exotic dark matter models \citep{2018Natur.555...71B,2018PhRvL.121a1101F,2018Natur.557..684M,2018PhRvD..98b3501L}, superconducting cosmic strings \citep{2019JCAP...09..009B}, or primordial black holes \citep{2018PhRvD..98b3503H,2018PhRvD..98d3006C,2022JCAP...03..030M} have all been proposed to explain the signal, while conventional astrophysical models have invoked an enhanced radio background sources from galaxies \citep{2018ApJ...858L..17F,2018ApJ...868...63E,2019MNRAS.483.1980M,2020MNRAS.499.5993R}. Meanwhile, a number of experiments are underway to verify the tentative detection\footnote{In addition to observational attempts, several papers have suggested potential instrumental systematics as the cause for the deep, flattened Gaussian signal \citep[e.g.][]{2018Natur.564E..32H,2019ApJ...880...26S,2019ApJ...874..153B,2020MNRAS.492...22S}.} such as SARAS \citep[][which has refuted the EDGES observation based on independent observational data at $2\sigma$ confidence]{2022NatAs...6..607S}, REACH \citep{2022NatAs...6.1332D}, MIST \citep{2024MNRAS.530.4125M}, RHINO \citep{2024arXiv241000076B}, PRIZM \citep{2019JAI.....850004P} and others\footnote{See \href{https://github.com/JitenDhandha/21cmExperiments}{github.com/JitenDhandha/21cmExperiments} for a comprehensive list of 21-cm experiments.}, with many already deployed and collecting data. Alongside this, measurements of the 21-cm signal spatial fluctuations (power spectrum) are being attempted by a number of interferometric arrays. In particular, the strongest limits so far from recent observations by HERA \citep{2023ApJ...945..124H}, LOFAR \citep{2025A&A...698A.186M,2025A&A...699A.109G} and MWA \citep{2025ApJ...989...57N} have been able to place the first limits on the presence of a heated IGM at $z \lesssim 11$, whilst other observations continue to push the high-redshift frontier \citep[e.g. NenuFAR;][]{2025MNRAS.542.2785M}.

Meanwhile, the field of high-redshift galaxy formation has also seen a rapid growth, with the advent of the JWST \citep{2006SSRv..123..485G} and its deep surveys. These surveys have revealed a wealth of information about the early Universe, including an abundance of bright galaxies at redshifts $z\gtrsim 10$ \citep[e.g.][]{2023ApJ...948L..14C,2023ApJ...952L...7A,2023ApJ...954L..46L,2023ApJS..265....5H,2024ApJ...964...71H,2024ApJ...970...31R,2024ApJ...969L...2F,2025ApJ...992...63W,2025ApJ...991..179P}, beyond what is expected from extrapolations of the UV luminosity functions (UVLFs) at lower redshifts. In order to resolve this apparent discrepancy, modifications to standard galaxy-formation models have been suggested, including a top-heavy initial mass function of early stars \citep[e.g.][]{2022ApJ...938L..10I,2023ApJ...954L..48W,2024MNRAS.529.3563T,2025A&A...694A.254H}, an increased SFE \citep[e.g.][]{2023MNRAS.523.3201D,2025A&A...696A.157M,2025MNRAS.538.3210B,2025MNRAS.542.2292D,2025arXiv250505442S},  or star-formation variability/stochasticity \citep[e.g.][]{2023MNRAS.521..497M,2023MNRAS.519..843M,2023ApJ...955L..35S,2024arXiv240504578K}.

As we enter the era of next-generation telescopes with the Square Kilometer Array \citep[][aiming to directly probe neutral hydrogen during the EoR and Cosmic Dawn]{2015aska.confE...1K} and the Athena X-ray Observatory \citep[][aiming to probe hot gas and X-ray populations at high-$z$]{2013arXiv1306.2307N}, the synergy between multi-wavelength observations and theoretical models is becoming increasingly important. Alongside this, advancements in Bayesian inference methods have made it possible to perform multi-dimensional, statistically rigorous inference of astrophysical parameters \citep[e.g.][]{2023MNRAS.525.6097S,2023MNRAS.524.4239P,2024A&A...688A.199P}, especially combining multi-wavelength data \citep[e.g.][]{2019MNRAS.483.1980M,2020MNRAS.495..123Q,2021MNRAS.507.2405C,2023ApJ...959...49D,2024MNRAS.527.9833B,2024MNRAS.531.1113P,2025JCAP...10..047K,2025arXiv250409725S}. In this work, we build on the simulations and Bayesian analysis performed in \citet{2025MNRAS.542.2292D}, where we placed strong constraints on key astrophysical parameters of the early Universe such as the SFE of Pop~II galaxies, and X-ray and radio emission efficiency of galaxies at $z\gtrsim 6$. We extend the analysis to infer the kinetic temperature and radio background temperature of the neutral IGM, and derive stringent limits on the 21-cm global signal and power spectrum within the context of our astrophysical model.

The paper is organized as follows: in Section~\ref{sec:method}, we describe the astrophysical model used in our simulations (Section~\ref{sec:astro_model}), the five observables of interest that we calculate in the simulations (Section~\ref{sec:observables}), the multi-wavelength observational data used to constrain the models (Section~\ref{sec:observational_datasets}), the Bayesian analysis framework used for inference (Section~\ref{sec:joint_analysis}), and a summary of the astrophysical parameter constraints from \citet[][Section~\ref{sec:astro_constraints}]{2025MNRAS.542.2292D}. In Section~\ref{sec:results}, we present the results of our analysis, including the neutral IGM temperature (Section~\ref{sec:derived_constraints}), and the 21-cm signal (Section~\ref{sec:21cm_signal_constraints}). We discuss the model dependency and caveats of our work in Section~\ref{sec:model_dependency}, and finally conclude in Section~\ref{sec:conclusions}. 

\section{Methodology} \label{sec:method}

\subsection{Astrophysical model}
\label{sec:astro_model}

We use the semi-numerical code \simcode\ \citep{2012Natur.487...70V,2012MNRAS.424.1335F,2013MNRAS.432.2909F,2014Natur.506..197F}\footnote{For a description of the code and relevant development papers, see \href{https://www.cosmicdawnlab.com/21cmSPACE/}{cosmicdawnlab.com/21cmSPACE}.} to generate 30,000 cosmological simulations of the early Universe. The simulations are described in detail in \citet{2025MNRAS.542.2292D}, but we summarize the key aspects here. The simulations consist of periodic boxes of cosmological volume $(128 \times \SI{3}{cMpc})^3$, spanning the redshift range $z=50$ to $z=6$. Large-scale overdensity fields $\delta$ and baryon-dark matter streaming velocity $v_\text{bc}$ \citep{2010PhRvD..82h3520T} are initialized using \textsc{Camb} \citep{2011ascl.soft02026L} and evolved linearly through time, while the gas kinetic temperature $T_\text{K}$ and ionized fraction $x_\text{e}$ are initialized using \textsc{Recfast} \citep{2011ascl.soft06026S} and evaluated at each time-step. Dark matter (DM) halos are modelled using the hybrid halo mass function $dn(M_h,z|\delta,v_\text{bc})/dM_h$ (HMF) of \citet{2011MNRAS.418..906T} \citep[see also][]{2004ApJ...609..474B}. Star-formation takes place in these halos above the cooling threshold quantified by $M_\text{crit}$, which is affected by the minimum virial circular velocity for star-formation $V_c$ (free parameter), as well as $v_\text{bc}$ suppression \citep{2012MNRAS.424.1335F,2022MNRAS.511.3657M}, inhomogeneous Lyman-Werner feedback \citep{2013MNRAS.432.2909F,2022MNRAS.511.3657M} and photoheating feedback \citep{2013MNRAS.432.3340S,2016MNRAS.459L..90C}. Star-formation begins with a single burst of Pop~III stars as halos cross $M_\text{crit}$ with an efficiency $f_\mathrm{\star,III}$, followed by a period of recovery due to supernovae feedback $t_\text{delay}$ \citep{2022MNRAS.514.4433M} that delays gas collapse into Pop~II stars. As we focus on Pop~II modelling in this work and \citet[][but see Section~\ref{sec:pop3_caveat} for more on this caveat]{2025MNRAS.542.2292D}, we fix these parameters to $f_\mathrm{\star,III}=0.2\%$ and $t_\text{delay}=\SI{30}{Myr}$, further assuming a log-flat IMF in the mass range $\SIrange{2}{180}{M_\odot}$ \citep{2022MNRAS.516..841G,2025NatAs...9.1268G}. 

Pop~II star-formation proceeds with a halo-mass and redshift dependent efficiency introduced in \citet{2025MNRAS.542.2292D}:
\begin{equation}
f_{\star,\twoi}(\mathbf{x},M_h,z) = 
\tilde{f}_{\star,\twoi} \begin{cases}
0 & M_h < M_\text{crit} \\
\dfrac{\log\left(M_h/M_\text{crit}\right)}{\log\left(M_\text{atm}/M_\text{crit}\right)} & M_\text{crit}\leq M_h < M_\text{atm} \\
1 & M_\text{atm}\leq M_h < M_\text{high}  \\
\left(\dfrac{M_h}{M_\text{high}}\right)^{\alpha_\star} & M_\text{high} \leq M_h
\end{cases}
\label{eqn:sfe_model}
\end{equation}
\noindent where $M_\text{atm}\propto(1+z)^{-3/2}$ is the atomic cooling threshold, and the turning point $M_\text{high}(z)$ is defined as:
\begin{equation}
    M_\text{high}(z) = M_0 \left(\dfrac{1+z}{7}\right)^{-\beta_\star}.
    \label{eqn:mhigh}
\end{equation}
The turning point determines the transition between a constant SFE of low-mass halos and a power-law SFE of high-mass halos, and $\beta_\star>0$ captures an enhancement of the stellar mass fraction at high-$z$ for a fixed halo mass $M_h$ \citep[see Figure~1 of][]{2025MNRAS.542.2292D}. The Pop~II star-formation is averaged over a time-scale of $t_\mathrm{\star,\twoi}=0.2H(z)^{-1}$, where $H(z)^{-1}$ is the Hubble time at redshift $z$ \citep[Equation 1 in][]{2025MNRAS.542.2292D} to get the Pop~II star-formation rate per halo ($\dot{M}_{\star,\twoi}$), or the HMF-integrated star-formation rate density $\dot{\rho}_\mathrm{\star,\twoi}(\mathbf{x},z)$ in each cell. The fully flexible Pop~II SFE is thus governed by four free parameters: $\tilde{f}_{\star,\twoi}$, $M_0$, $\alpha_\star$ and $\beta_\star$.

In order to model the radiative transfer from stellar/galactic sources, we calculate the specific emissivity $\epsilon(\mathbf{x},\nu,z)$ in the Lyman band \citep[responsible for the Ly$\alpha$ coupling of the 21-cm signal and Ly$\alpha$ heating; e.g.][]{2021MNRAS.506.5479R}, as well as X-ray and radio bands (described below). They are propagated throughout the simulation via window functions. Of particular interest in this work are the X-ray emissions, likely sourced by low metallicity high mass X-ray binaries at high redshifts \citep[HMXBs; e.g.][]{2013ApJ...776L..31F,2016ApJ...825....7L}. We assume the X-ray luminosity per unit energy is proportional to the halo SFR as in local star-burst galaxies \citep[e.g.][]{2003MNRAS.339..793G,2010ApJ...724..559L,2012MNRAS.419.2095M,2014MNRAS.437.1698M}\footnote{The bolometric luminosity per star-formation rate ($L_X/\text{SFR}$) often used in literature is just the integrated luminosity over the X-ray SED, i.e. $\int \hat{L}_X(E)/\text{SFR}\, dE$.}:
\begin{equation}
    \dfrac{\hat{L}_X(E)}{\mathrm{SFR}} = f_\text{X} \times \left(3\times 10^{40}
    \SI{}{erg~s^{-1}~M_\odot^{-1}~yr}\right)\hat{\epsilon}_X(E)
    \label{fX_equation}
\end{equation}
where $\hat{\epsilon}_X(E)$ is the X-ray SED (normalized to 1) from \citet[][the mean model at $z=15.34$ in the energy range $\SIrange{0.2}{95}{keV}$ including interstellar absorption]{2013ApJ...776L..31F} in units of $\SI{}{eV}^{-1}$. The luminosity normalization is chosen such that it matches the theoretical prediction for low metallicity HMXBs \citep{2013ApJ...764...41F,2013ApJ...776L..31F,2021ApJ...907...17L}. We further assume that Pop~II and Pop~III X-ray binaries have the same normalization, i.e. $f_\text{X}=f_{\text{X},\twoi}=f_{\text{X},\threei}$ (varied as a free-parameter), as well as the same SED. Thus, the comoving X-ray emissivity can be calculated as: $\epsilon_X(\mathbf{x},E,z) = \left(\hat{L}_X(E)/\text{SFR}\right)\times\dot{\rho}_{\star,\twoi+\threei}(\mathbf{x},z)$. The X-rays contribute to both heating, as well as ionization of the IGM \citep[e.g.][]{1996ApJ...472L..63O,2001ApJ...563....1V,2007MNRAS.376.1680P,2012ApJ...760....3M}.

Similar to the X-ray emissivity, we model the radio luminosity per unit frequency of galaxies at high-$z$ as \citep{2018MNRAS.475.3010G, 2019MNRAS.483.1980M, 2020MNRAS.499.5993R}:
\begin{equation}
    \dfrac{L_r(\nu)}{\text{SFR}} = f_\text{r} \times \left(\SI{10^{22}}{W~s^{-1}~Hz^{-1}~M_\odot^{-1}~yr}\right)\left(\frac{\nu}{\SI{150}{MHz}}\right)^{-0.7},
\label{fr_equation}
\end{equation}
where the spectral index is typical for radio synchrotron emissions, and the normalization is with respect to present-day radio galaxies \citep{2016MNRAS.462.1910H, 2018MNRAS.475.3010G}. Like X-rays, we assume the same emission for both Pop~II and Pop~III stars, i.e. $f_\text{r}=f_{\text{r},\twoi}=f_{\text{r},\threei}$ (and vary it as a free parameter).

The emissivities thus calculated are then used to determine the heating and ionization rates of the IGM. The evolution of the kinetic/matter temperature $T_\text{K}$, is governed by the balance between heating and cooling processes \citep[e.g.][]{2006PhR...433..181F,2011MNRAS.411..955M}:
\begin{align}
\frac{dT_\text{K}}{dz} &= \frac{2 T_\text{K}}{1 + z} + \frac{2 T_\text{K}}{3(1 + \delta)} \frac{d\delta}{dz} - \frac{d x_\text{e}}{dz} \cdot \frac{T_\text{K}}{1 + x_e} \\
&\quad - \frac{2}{3 k_B (1 + f_{\text{He}} + x_\text{e})} \left( Q_X + Q_{\text{Compton}} + Q_{\text{Ly}\alpha}\right),
\end{align}
where $f_\text{He}$ is the helium fraction, and $x_\text{e}$ is the free electron fraction. In the above equation, the terms (from left to right) account for cooling due to Hubble/adiabatic expansion and structure formation, and heating due to change in number of ionized particles, X-rays ($Q_X$), Compton scattering ($Q_{\text{Compton}}$) and Ly$\alpha$ photons \citep[$Q_{\text{Ly}\alpha}$, accounting for multiple scattering effects,][]{2021MNRAS.506.5479R}. 

We compute the ionization by UV photons from galaxies and X-rays using the excursion set formalism commonly used in semi-numerical simulations \citep{2004ApJ...613....1F}. A region of the simulation is considered to be ionized if it satisfies
\begin{align}
    \exists R<R_\text{mfp}\ &,\quad {\rm s.t.}\quad \zeta_\text{ion} f_\text{coll}(\mathbf{x},R) > 1 - x_\text{e,oth}(\mathbf{x},R).
    \label{eqn:reionization}
\end{align}
Here, $\zeta_\text{ion}$ is the ionization rate per baryon (varied as a free parameter), $R_\text{mfp}$ is the maximum mean free path of ionizing photons \citep{2015MNRAS.449.4246G,2017MNRAS.472.1915C} set to $\SI{50}{cMpc}$ \citep[e.g.][]{2004Natur.432..194W,2021MNRAS.506.2390Q,2023ApJ...955..115Z}, and $f_\text{coll}(\mathbf{x},R)$ is the fraction of baryons collapsed into galaxies averaged over a volume of radius $R$ centered at $\mathbf{x}$. If the above criterion is not satisfied, the cell is considered to have a two-phase IGM with partial ionization. The ionizations from long range agents like X-rays are encapsulated in the final term:
\begin{equation}
    \dfrac{dx_\text{e,oth}(\mathbf{x},z)}{dz} = \dfrac{dt}{dz}\left[\Lambda_\text{ion,X} - \alpha_\text{B} n_\text{H}(\mathbf{x}) x_\text{e,oth}^2
    \right],
\end{equation}
where $\Lambda_\text{ion,X}$ is the X-ray ionization rate per baryon, $\alpha_\text{B}$ is the case-B recombination coefficient, and $n_\text{H}(\mathbf{x})$ is the large-scale hydrogen number density. The neutral fraction is then given by ${x_\text{HI}(\mathbf{x}) = 1-x_\text{e} = 1 - \zeta_\text{ion} f_\text{coll}(\mathbf{x}) - x_\text{e,oth}(\mathbf{x})}$.

The astrophysical model described in this section thus consists of eight free parameters, listed in Table~\ref{priortable}. The priors are chosen to be broad enough to cover the range of astrophysical models that can be expected at high-$z$, while also being physically motivated (e.g. a $V_c$ of $\SI{4.2}{km~s^{-1}}$ corresponds to the molecular cooling threshold). Furthermore, in place of $\zeta_\text{ion}$, we use $\tau_\text{CMB}$ (which is calculated in post processing) to quantify the reionization history and directly sample this derived parameter in our Bayesian analysis \citep[following previous works, e.g.][]{2024MNRAS.527..813B,2024MNRAS.531.1113P}.

\begin{table}
\renewcommand\arraystretch{1.1}
 \caption{\label{priortable}Astrophysical parameter priors used in \citet{2025MNRAS.542.2292D} and this work. The parameters $\{V_c, \tilde{f}_{\star,\twoi}, M_0, \alpha_\star, \beta_\star\}$ described the star-forming properties of early galaxies, while $\{\tau_\text{CMB}, f_\text{X}, f_\text{r}\}$ describe their radiative properties (in particular, their ionizing, X-ray and radio emission efficiencies). We use the derived parameter $\tau_\text{CMB}$ instead of $\zeta_\text{ion}$ to quantify the reionization history, allowing us to use \textit{Planck 2018} constraints on the optical depth as our prior, and allow for a wide prior to remain agnostic to early/late reionization scenarios \citep[e.g.][]{2024MNRAS.533.2843A,2025arXiv250821069E}.}
 \label{tab:astro_priors}
 \centering
 \begin{tabular*}{1\linewidth}{@{\extracolsep{\fill}}cccc}
  \hline
  \textbf{Parameter} & \textbf{Prior} & \textbf{Minimum} & \textbf{Maximum} \\
  \hline
  $V_c$ & Log-Uniform & $\SI{4.2}{km~s^{-1}}$ & $\SI{100}{km~s^{-1}}$ \\
  $\tilde{f}_{\star,\twoi}$ & Log-Uniform & $10^{-4}$ & $10^{-0.3}$  \\
  $M_0$ & Log-Uniform & $\SI{2\times10^8}{M_\odot}$ & $\SI{10^{11}}{M_\odot}$ \\
  $\alpha_\star$ & Uniform & $0$ & $2$ \\
  $\beta_\star$ & Uniform & $0$ & $5$ \\
  $\tau_\text{CMB}$ & Uniform & $0.033$ & $0.075$ \\
  $f_\text{X}$ & Log-Uniform & $10^{-3}$ & $10^3$  \\
  $f_\text{r}$ & Log-Uniform & $10^{-1}$ & $10^5$ \\
   \hline
 \end{tabular*}
\end{table}

\subsection{Observables of interest}
\label{sec:observables}

The 21-cm signal is sensitive to the astrophysical processes described above, and can be quantified through the sky-averaged global signal $T_{21}(z)$, or the power spectrum $\Delta_{21}^2(k,z)$. The strength of the signal is given by the differential brightness temperature \citep[e.g.][]{1997ApJ...475..429M,2006PhR...433..181F,2012RPPh...75h6901P}:
\begin{equation}
    T_{21}(\mathbf{x},z) = \left(1-e^{-\tau_{21}(\mathbf{x},z)}\right)\dfrac{T_\text{S}(\mathbf{x},z)-T_\text{rad}(\mathbf{x},z)}{1+z},
    \label{eqn:T21_equation}
\end{equation}
where $\tau_{21}$ is the 21-cm radiation optical depth, $T_\text{S}$ is the spin temperature, and $T_\text{rad}$ is the radio background temperature at the rest-frame 21-cm wavelength at redshift $z$. In regions of the Universe where $T_\text{S}>T_\text{rad}$, the signal is seen in emission (and in absorption otherwise). The spin temperature $T_\text{S}$ is given by the balance between collisional and radiative processes, and is defined as:
\begin{equation}
    T_\text{S}(z)^{-1} = \dfrac{x_\text{rad}T_\text{rad}^{-1} + x_\alpha T_\text{C}^{-1} + x_\text{c} T_\text{K}^{-1}}{x_\text{rad} + x_\alpha + x_\text{c}},
    \label{eqn:TS_equation}
\end{equation}
where we have suppressed the spatial dependence in the equation for brevity. The terms $x_\text{rad}$\footnote{We assume $x_\text{rad}=(1-e^{-\tau_{21}})/\tau_{21}\approx 1$ in this work, which is valid in the optically thin limit $\tau_{21}\rightarrow 0$ at the redshifts of interest.}, $x_\alpha$ and $x_\text{c}$ are the relative coupling strengths to the CMB, Ly$\alpha$ background via the Wouthuysen-Field coupling mechanism \citep{1952AJ.....57R..31W,1958PIRE...46..240F}, and collisional processes respectively \citep[e.g.][]{1956ApJ...124..542P,2005ApJ...622.1356Z}.

Averaging over the sky at a particular frequency (or in case of simulations, the brightness temperature field at a particular redshift) gives us the global signal:
\begin{equation}
    T_{21}(z) = \langle T_{21}(\mathbf{x},z) \rangle_\mathbf{x}.
    \label{eqn:T21_global}
\end{equation}
On the other hand, the power spectrum $P_{21}(k,z)$, which quantifies the spatial fluctuations of the brightness temperature field, is defined as:
\begin{equation}
    \bigl\langle \tilde{T}_{21}(\mathbf{k},z) \tilde{T}_{21}^*(\mathbf{k'},z) \bigr\rangle = (2\pi)^3 \delta^D(\mathbf{k}-\mathbf{k'})P_{21}(k,z).
\end{equation}
where $\tilde{T}_{21}(\mathbf{k},z)$ is the Fourier transform of $T_{21}(\mathbf{x},z)$. For consistency with other works, we define the dimensionless power spectrum:
\begin{equation}
\Delta_{21}^2(k,z) = (k^3/2\pi^2)P_{21}(k,z)
\end{equation}
to quantify the spatial fluctuations of the 21-cm signal in this work.

Due to the versatile nature of the 21-cm signal and its sensitivity to the astrophysical processes described above, it can be used in conjunction with other observables of the early Universe. The diffuse X-ray background and radio background seen today are complementary probes setting upper limits on radiative efficiencies of galaxies at high-$z$ \citep{2017MNRAS.464.3498F,2025ApJ...983...54M}. Following previous works \citep[e.g.][]{2024MNRAS.531.1113P,2025MNRAS.542.2292D}, we define the present-day angle-averaged X-ray specific intensity from sources at $z>6$ as \citep[e.g.][]{2007MNRAS.376.1680P}:
\begin{equation}
    J_X(E,z=0) = \frac{1}{4 \pi} \int_{z'=6}^{\infty} \epsilon_X(E',z')e^{- \tau_X(E',z')} \bigg|\dfrac{cdt}{dz'}\bigg|dz',
\label{JX_equation}
\end{equation}
where $E'=E(1+z')$ is the energy at emission redshift $z'$, $\epsilon_X(E',z')$ is the comoving X-ray emissivity defined in the previous section (but averaged across the simulation box), and $\tau_X(E',z')$ is the X-ray optical depth. We can further define the integrated intensity in an energy band $E_\text{band}$ as:
\begin{equation}S_X\left(E_\text{band}\right)=\int_{E_\text{band}}J_X(E,z=0)dE
    \label{eqn:SX_equation}
\end{equation}
for direct comparison with broad-band Chandra (and other) X-ray observations. Similarly, we define the radio background temperature seen today as \citep[e.g.][]{2020MNRAS.492.6086E,2020MNRAS.499.5993R}\footnote{Note that in the absence of an explicit $\nu$ variable, $T_\text{rad} = T_\text{rad}(\nu_{21},z)$, the radio temperature at the rest-frame 21-cm wavelength at a given redshift, as used in Equation~\ref{eqn:T21_equation} and for the rest of the paper.
}:
\begin{equation}
    T_\text{rad}(\nu,z = 0) = T_\text{CMB} + 
    \frac{c^2}{2k_{\rm B} \nu^2} \dfrac{1}{4\pi}
    \int_{z'=6}^{\infty} \epsilon_r(\nu',z')\bigg|\dfrac{cdt}{dz'}\bigg|dz',
    \label{Tr_equation}
\end{equation}
where the second term accounts for the present-day angle-averaged radio specific intensity from sources at $z'>6$, and $\nu'=(1+z')$ is the frequency at emission redshift $z'$, and radio photons are assumed to be free-streaming (i.e., negligible absorption or scattering since the optical depth is $\ll 1$). Observations of radio excess at $<\SI{10}{GHz}$ \citep[e.g.][]{2011ApJ...734....5F,2017MNRAS.469.4537D} can help constrain the radio emissivity of galaxies at high-$z$, by setting upper limits on their contribution to the radio background observed today.

In \citet{2025MNRAS.542.2292D}, UVLFs were introduced as a new output in \simcode\ to derive constrain on the SFE of high-redshift galaxies using HST and JWST observations. The UVLF measures the number density of galaxies per unit magnitude in the ultraviolet band, defined as $\lambda_\text{UV} \equiv \SI{1500}{\angstrom} \pm \SI{50}{\angstrom}$. In order to model the UVLF, we employ a framework that uses the hybrid HMF described in Section~\ref{sec:astro_model} \citep[e.g.][]{2019MNRAS.484..933P}:
\begin{equation}
    \Phi(\mathbf{x},M_\text{UV},z) = \big(1+\delta(\mathbf{x})\big)\dfrac{dn(\mathbf{x},M_h,z)}{dM_h}\left|\dfrac{M_h}{M_\text{UV}}\right|,
    \label{eqn:uvlf_def}
\end{equation}
where the $(1+\delta)$ factor accounts for the conversion of the HMF from Lagrangian space to Eulerian space \citep[see, e.g.][]{2011MNRAS.411..955M,2011MNRAS.418..906T,2023MNRAS.523.2587M}. To evaluate the final term in the above equation, we map the host halo mass $M_h$ to the UV magnitude $M_\text{UV}$ of the galaxy contained within, using the SFR$-$UV relation:
\begin{equation}
    \dot{M}_{\star,\twoi}(\mathbf{x},M_h,z) = \kappa_\text{UV} \times L_\text{UV}(\mathbf{x},M_h,z),
    \label{eqn:SFR_UV_equation}
\end{equation}
where $\kappa_\text{UV}=\SI{1.15\times10^{-28}}{M_\odot~yr^{-1}~erg^{-1}~s^{-1}~Hz}$ \citep[][]{2014ARA&A..52..415M}. This value assumes a continuous star-formation history and a Salpeter IMF spanning the mass range $\SIrange{1}{100}{M_\odot}$. In the early Universe, the UV luminosity may be dominated by a stellar population that is strongly metallicity dependent or top-heavy in stellar mass \citep[e.g.][]{2023ARA&A..61...65K}. We use the above $\kappa_\text{UV}$ for consistency with other works in the literature, and leave a self-consistent, spatially varying and time-evolving Pop~II spectra calculation to future work. However, we do include an analytic dust model to correct for the attenuation of the UV luminosity, which is expected to be crucial at $z\lesssim10$ where HST data is strongest. The dust distribution across galaxies can be quantified using their UV spectral slope $\beta$ as a proxy \citep[e.g. infrared excess and $\beta$ relation of][]{1999ApJ...521...64M}, which depends on the intrinsic UV luminosity of the galaxy as well as the redshift. Here, we use the $\beta(M_\text{UV}^\text{dust},z)$ fit from \citet{2024JCAP...09..018Z} which combines observations from HST \citep{2014ApJ...793..115B} and JWST \citep{2024MNRAS.529.4087T}, making it applicable across a wide redshift range. The dust attenuation can thus be thought as mapping $M_\text{UV}^\text{dust-free}(z) \rightarrow M_\text{UV}^\text{dust}(M_\text{UV}^\text{dust-free},z)$. In this calibration, the attenuation is negligible at $z\gtrsim10$ and for faint galaxies that are insufficiently dust-rich. The above calculations are performed in each pixel to get the UVLF at $\SI{3}{cMpc}$ scale, and then averaged across the simulation volume to capture the statistics of large-scale surveys. 

\subsection{Observational datasets}
\label{sec:observational_datasets}

Having computed the astrophysical observables of interest, we now describe the data used in our joint analysis. We use five observational datasets, following the analysis in \citet{2024MNRAS.531.1113P} and \citet{2025MNRAS.542.2292D}, as summarized below.
\begin{itemize}
    \item \textbf{21-cm global signal non detection} from SARAS~3: We use global sky temperature measurements from SARAS~3 \citep{2021ExA....51..193T,2022NatAs...6..607S} to constrain the global 21-cm signal directly, in the redshift range $z\approx\SIrange{15}{25}{}$ (or $\nu \approx\SIrange{55}{85}{MHz}$). The data consists of 15 hours of observations from its deployment in southern India, and is calibrated to remove any radio-frequency interference, receiver systematics, and environmental thermal emissions. We fit the data with a 6th order log-log polynomial as the Galactic and extragalactic foreground model:
    \begin{equation}
        \log_{10}\left(T_\text{fg}(\nu)/\text{K}\right) = \sum_{i=0}^{6} a_i \left[f_\text{N}\left(\log_{10}\left(\nu/\text{MHz}\right)\right)\right]^i
    \end{equation}
    and a thermal noise term as done in previous works \citep[e.g.][]{2022NatAs...6..607S,2022NatAs...6.1473B,2024MNRAS.527..813B}. In the above equation, $f_\text{N}$ is a function that normalizes the log-frequency to the range $[-1,1]$. We marginalize over the foreground + noise parameters in our Bayesian analysis \citep[e.g.][]{2024MNRAS.527..813B,2024MNRAS.531.1113P,2024MNRAS.529..519G} when discussing constraints on the astrophysical parameters. The residuals, although quite sensitive to the foreground fit, are generally of the order of $\sim\SIrange{200}{300}{mK}$. 
    
    \item \textbf{21-cm power spectrum upper limits} from HERA Phase 1: In addition to the global signal,
    we use 21-cm power spectrum limits from HERA Phase 1 observations \citep{2023ApJ...945..124H}. These are the most stringent upper limits to date --- using data from 94 nights of observations with $\SIrange{34}{41}{}$ antennas --- at redshifts $z\approx7.9$ and $z\approx10.4$ in the wavevector range $k\approx\SIrange{0.2}{0.4h}{Mpc}^{-1}$. We use the frequency bands least contaminated by RFI (Band 1 in Field D, and Band 2 in Field C) and decimate neighbouring $k$-bins to remove correlations. Since foreground avoidance is employed, only residual systematics need to be marginalised over, which we do using the likelihood function from \citet{2023ApJ...945..124H}.

    \item \textbf{Cosmic X-ray background (CXB)}: The diffuse X-ray background is expected to contain contributions from unresolved faint point sources such as AGN and galaxies, diffuse emission from the hot IGM, as well as redshifted emission from unresolved X-ray binaries in the early Universe --- the latter being the focus of this study. We use CXB flux measurements from Chandra \citep{2006ApJ...645...95H} and the collated data from Table 1 of \citet{2016ApJ...831..185H}. This includes measurements from HEAO \citep{1980ApJ...235....4M,1999ApJ...520..124G}, BeppoSAX \citep{2007ApJ...666...86F}, INTEGRAL \citep{2007A&A...467..529C} and Swift BAT \citep{2008ApJ...689..666A}. 

    \item \textbf{Cosmic radio background (CRB)}: Similar to the CXB, the unresolved diffuse radio background (after removal of Galactic foregrounds and CMB) is expected to contain contributions from unresolved radio galaxies at low and high-$z$. We use the CRB data collated in Table 2 of \citet{2018ApJ...858L...9D}, which includes measurements from LWA1 Low Frequency Sky Survey \citep[$\SIrange{40}{80}{MHz}$;][]{2017MNRAS.469.4537D}, ARACDE2 \citep[$\SIrange{3}{11}{GHz}$;][]{2011ApJ...734....5F}, and other single frequency experiments at $\SI{22}{MHz}$ \citep{1999A&AS..137....7R}, $\SI{46}{MHz}$ \citep{1997A&AS..124..315A,1999A&AS..140..145M}, $\SI{408}{MHz}$ \citep{1982A&AS...47....1H,2015MNRAS.451.4311R}, and $\SI{1420}{MHz}$ \citep{2001A&A...376..861R}.

    \item \textbf{UV luminosity functions (UVLF)} from HST/JWST: We use multi-field HST derived UVLF determinations from \citet[][see their Table 4]{2021AJ....162...47B} in the redshift range $z=6-9$, and JWST derived UVLFs from \citet[][see their Table 2]{2024MNRAS.533.3222D} in the redshift range $z=9-14.5$. The former contains 1946 sources from HUDF/XDF and parallel fields, BORG/HIPPIES, five CANDELS fields, HFF and parallel fields, and observations from the ERS program (see their Table~1). The latter contains 2548 sources from PRIMER, NGDEEP and JADES fields. We do not use more datasets at high-$z$ to minimize source-overlap (which would artificially shrink error bars) and to avoid systematic differences between analysis pipelines. Nonetheless, \citet{2024MNRAS.533.3222D} is broadly consistent with other JWST-based UVLF measurements \citep[e.g.][]{2024MNRAS.527.5004M,2023ApJ...954L..46L,2023ApJ...951L...1P,2024ApJ...965..169A,2024ApJ...970...31R}, lending further confidence to their findings and our results.
    
    \item \textbf{CMB optical depth} ($\tau_\text{CMB}$) from \textit{Planck 2018}: We indirectly use the CMB optical depth as a constraint by adopting a uniform prior on the reionization history of $3\sigma$ around the measured \textit{Planck 2018} value of $\tau_\text{CMB}=0.054\pm0.07$ \citep{2020A&A...641A...6P}, i.e. $\tau_\text{CMB} \in \left[0.033,0.075\right]$ as shown in Table~\ref{tab:astro_priors}. This ensures that our reionization histories are not too extreme, while also allowing for a wide range of astrophysical models.
    
\end{itemize}

\subsection{Joint Bayesian analysis}
\label{sec:joint_analysis}

\begin{table*}
\centering
\caption{Summary of the datasets used in \citet{2025MNRAS.542.2292D} and this work, and their different combinations for joint analysis. The first five rows show individual datasets (in case of CXB and CRB, the UVLF is included as explained in Section~\ref{sec:joint_analysis}), while the latter four rows show the joint analysis cases. The `Joint w/o UVLF' is similar to the analysis in \citet{2024MNRAS.531.1113P}, although we use a different SFE model and exclude Pop~III free parameters. This case is included to show the importance of the UVLFs in early Universe constraints, as demonstrated in \citet{2025MNRAS.542.2292D}. The `Joint w/o high-$z$ UVLF' case is useful to assess how our knowledge of the 21-cm signal has changed with the advent of JWST observations (i.e. how HST-inferred fixed SFE models change our conclusions). Finally, since the SARAS~3 data includes nuisance parameters to subtract the radio foregrounds (that we marginalize over in our results), we also show a `Joint w/o SARAS~3' case to allow for the case that the foreground fit is inaccurate.}
\resizebox{1\linewidth}{!}{
\begin{threeparttable}
\begin{tabular}{l|c|c|c|c|c|c}
\hline
\textbf{Fit type} & \textbf{21-cm global signal} & \textbf{21-cm power spectra} & \multicolumn{2}{c|}{\textbf{UV luminosity functions}} & \textbf{Cosmic X-ray bg.} & \textbf{Cosmic radio bg.} \\
& (SARAS~3, \citeyear{2022NatAs...6..607S})
& (HERA, \citeyear{2023ApJ...945..124H})
& $z \leq 10$ \citepalias[HST;][]{2021AJ....162...47B}$^\dagger$
& $z>10$ \citepalias[JWST;][]{2024MNRAS.533.3222D}
& (Chandra + others) & (LWA1 + others) \\
\hline
UVLF & & & $\checkmark$ & $\checkmark$ & & \\
UVLF + CXB & & & $\checkmark$ & $\checkmark$ & $\checkmark$ & \\
UVLF + CRB & & & $\checkmark$ & $\checkmark$ & & $\checkmark$ \\
HERA & & $\checkmark$ & & & & \\
SARAS~3 & $\checkmark$ & & & & & \\
\hline
Joint w/o UVLF & $\checkmark$ & $\checkmark$ & & & $\checkmark$ & $\checkmark$ \\
Joint w/o high-$z$ UVLF & $\checkmark$ & $\checkmark$ & $\checkmark$ & & $\checkmark$ & $\checkmark$ \\
Joint w/o SARAS~3 & & $\checkmark$ & $\checkmark$ & $\checkmark$ & $\checkmark$ & $\checkmark$ \\
\textbf{Joint} & $\checkmark$ & $\checkmark$ & $\checkmark$& $\checkmark$ & $\checkmark$ & $\checkmark$ \\
\hline
\end{tabular}
\begin{tablenotes}
\item{$^\dagger$ Technically, the $z\leq10$ UVLFs contain both HST determinations from \citetalias{2021AJ....162...47B} at $z=\SIrange{6}{9}{}$ and JWST determinations from \citetalias{2024MNRAS.533.3222D} at $z=9,10$, but we refer to it as `HST' for simplicity. The data in this redshift range prefers a redshift-independent SFE, which is consistent with pre-JWST findings of constant SFE models.}
\end{tablenotes}
\end{threeparttable}
}
\label{tab:data_summary}
\end{table*}

\begin{figure*}
   \includegraphics[width=1\linewidth]{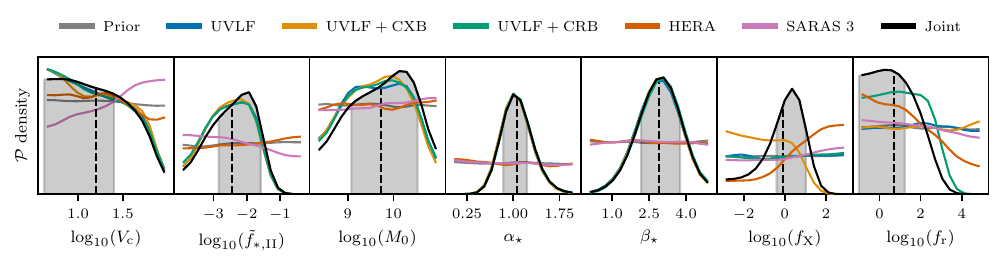}
    \caption{1D marginal posterior probability distributions (PDFs) of the seven astrophysical parameters (mariginalizing over $\tau_\text{CMB}$ since it remains prior-dominated) from the individual and joint analysis. The constraints shown here are the same as in \citet{2025MNRAS.542.2292D}, except for the case of CXB+UVLF and CRB+UVLF which are new to this work. In case of $\alpha_\star$ and $\beta_\star$, all datasets containing UVLF  lie on top of each other. The vertical black dashed line and grey shaded region show the weighted posterior mean and the 68\% credible interval (CI), respectively, for the joint analysis. For the special cases of the joint analysis, see Figure~\ref{fig:param_constraints_joint}.
    \label{fig:param_constraints}}
\end{figure*}

In order to derive the constraints on the astrophysical parameters described in Section~\ref{sec:astro_model} using our collection of observational datasets, we use the Bayesian nested-sampling package \polychord\ \citep{2015MNRAS.450L..61H,2015MNRAS.453.4384H}. Under the Bayesian statistical framework, we define a likelihood function for each observable such that $\mathcal{L(\theta)} = P(\mathcal{D}|\theta)$ is the probability of the data $\mathcal{D}$ given the parameters $\theta$. This can then be combined with a \textit{prior} probability distribution $\pi(\theta)$ to obtain the \textit{posterior} distribution $\mathcal{P}(\theta) = P(\theta|D)$:
\begin{equation}
    \mathcal{P}(\theta) = \dfrac{\mathcal{L}(\theta)\pi(\theta)}{\mathcal{Z}},
\end{equation}
where $\mathcal{Z} = P(\mathcal{D}) = \int \mathcal{L}(\theta)\pi(\theta) d\theta$ is the Bayesian evidence. The Bayesian inference allows us to marginalize over nuisance parameters (e.g. SARAS~3 foreground parameters), and obtain posterior distributions for the astrophysical parameters of interest. Futhermore, assuming statistical independence of the observables, we can write the joint likelihood as a product of all the individual likelihoods:
\begin{equation}
    \mathcal{L}_\text{Joint} = \mathcal{L}_\text{SARAS~3} \times\mathcal{L}_\text{HERA}\times \mathcal{L}_\text{CXB} \times \mathcal{L}_\text{CRB} \times \mathcal{L}_\text{UVLF}.
\end{equation}
or a subsets of these, depending on the analysis. We use upper limit likelihoods for the CXB, CRB and HERA datasets \citep[e.g. see Appendix~A of][]{2019ApJ...887..141L}, a Gaussian likelihood for SARAS~3 including a noise term $\sigma_\text{S3}$, and a two-piece Gaussian likelihood for the UVLF data to account for assymetric errors \citep[see][for more details]{2025MNRAS.542.2292D}. All likelihoods include uncertainties on the data, as well as on our model predictions (a $20\%$ error floor on simulation accuracy + additional emulator errors).

\subsection{Astrophysical constraints from \citet{2025MNRAS.542.2292D}}
\label{sec:astro_constraints}

We use the nested sampling runs from \citet{2025MNRAS.542.2292D} for the most part. They were performed by training neural network emulators on the observables of interest output by \simcode\footnote{Each simulation takes roughly $\sim\SIrange{2}{4}{hrs}$ while emulators can evaluate an observable on the order of $~\SIrange{1}{10}{ms}$ \citep[e.g.][]{2021MNRAS.508.2923B}.}, and then efficiently sampling the parameter space using \polychord. However, the case CXB and CRB datasets are different, since we also include UVLF data for both. The diffuse X-ray and radio backgrounds are sensitive to the SFR of galaxies (i.e., $\propto\text{SFR}\times f_\text{X}$ and $\text{SFR}\times f_\text{r}$), and the UVLF data breaks the degeneracy between the SFE, and $f_\text{X}$ or $f_\text{r}$, anchoring the former to observationally determined values. This yields more accurate constraints on $f_\text{X}$ and $f_\text{r}$ compared to previous works like \citet{2024MNRAS.531.1113P} and \citet{2024MNRAS.529..519G} that do not use UVLFs, and is one of the main results in \citealt{2025MNRAS.542.2292D} \citep[see also][who arrive at a similar conclusion]{2020MNRAS.495..123Q,2023ApJ...959...49D}.

Table~\ref{tab:data_summary} summarizes the combinations of datasets used in this work, and Figure~\ref{fig:param_constraints} shows the corresponding 1D marginal posterior probability distributions (PDFs) for the minimum virial circular velocity for star-formation in DM halos $V_c$, SFE parameters $\tilde{f}_{\star,\twoi}$, $M_0$, $\alpha_\star$, $\beta_\star$, and the X-ray and radio emission efficiencies $f_\text{X}$ and $f_\text{r}$. A similar plot for the special cases of the joint analysis is shown in Figure~\ref{fig:param_constraints_joint}, to assess how sensitive our constraints are to the inclusion of JWST data or to the SARAS~3 foreground fit. Since we use the joint constraints from \citet{2025MNRAS.542.2292D} to derive IGM properties and the 21-cm signal, we summarize the key findings here and refer the reader to the aforementioned work for more details.

The minimum virial circular velocity for star-formation $V_c$ is constrained by a combination of SARAS~3 and UVLFs. The former has a weak preference for higher values of $V_c$ so as to delay WF coupling and the onset of the 21-cm absorption signal to outside the SARAS~3 band\footnote{This constraint is highly sensitive on the foreground model parameters, as we shall discuss in Section~\ref{sec:without_SARAS3}.}, while the latter sets strong upper limits on $V_c$ by direct detection of faint galaxies at $M_\text{UV}\lesssim-18$ by HST/JWST \citep[see, e.g.][]{2025JCAP...10..047K}. The SFE parameters on the other hand, $\tilde{f}_{\star,\twoi}$, $M_0$, $\alpha_\star$, and $\beta_\star$, are entirely constrained by UVLFs. The halo mass dependence of the SFE is inferred to be $\alpha_\star\approx 1.1$ \citep[consistent with $M_\star\propto M_h^2$;][]{2018ApJ...868...92T}, while the redshift evolution, given by $\left[(1+z)/7\right]^{-\beta_\star}$ is inferred to have a non-monotonic behaviour. The UVLFs show a preference for $\beta_\star \lesssim 1.4$ at $z\lesssim 10$ \citep[i.e., minimal evolution, consistent with pre-JWST HST-based inferences; e.g.][]{2015ApJ...803...34B,2018ApJ...868...92T,2018PASJ...70S..11H}, and $\beta_\star\approx 3.6$ \citep[enhanced SFE in the early Universe; e.g.][]{2023MNRAS.523.3201D,2025MNRAS.538.3210B,2025arXiv250505442S}. The joint analysis gives an intermediate evolution of $\beta_\star\approx 2.9$. 

The X-ray and radio radiative efficiencies $f_\text{X}$ and $f_\text{r}$ (defined in Equations~\ref{fX_equation} and \ref{fr_equation}) are constrained at the upper end of their respective priors by the upper limits on the diffuse CXB and CRB. HERA provides a secondary contribution to both constraints; it disfavours low values of $f_\text{X}$ by requiring efficient heating at $z\lesssim11$ to suppress the 21-cm signal, and disfavours high values of $f_\text{r}$ which would otherwise lead to excess radio buildup, deepening the signal. Figure~\ref{fig:param_constraints_joint} shows how the constraints on $f_\text{X}$ and $f_\text{r}$ are strengthened by the inclusion of the UVLF data\footnote{\label{foot:fXfr_robustness}Although the constraints on $f_\text{X}$ and $f_\text{r}$ depend on UVLFs, they are robust to the choice of including or excluding high-$z$ galaxies at $z\gtrsim10$. This is because the largest contribution to diffuse X-ray and radio backgrounds comes from galaxies at $z\sim6$ where the SFE (in particular $\beta_\star$) parameter is anchored (as defined in Equation~\ref{eqn:mhigh}). This is clear by the match of black and dashed light-blue curves in Figure~\ref{fig:param_constraints_joint}.}.

\begin{figure}
   \includegraphics[width=1\linewidth]{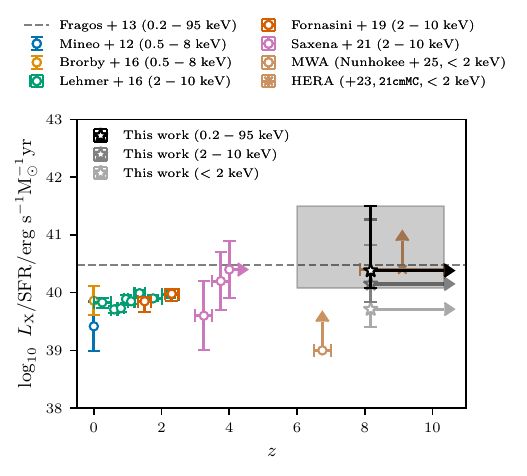}
    \caption{Constraints on the X-ray luminosity per unit SFR, $L_X/\text{SFR}$, from Chandra observations of star-forming galaxies in the local Universe \citep{2012MNRAS.419.2095M,2016MNRAS.457.4081B} and at low-redshifts \citep{2016ApJ...825....7L,2019ApJ...885...65F}, and 21-cm experiment inferred limits at high-redshifts \citep{2023ApJ...945..124H,2025ApJ...989...57N}. We put constraints at 
    $\log_{10}(L_X/\text{SFR}/\SI{}{erg~s^{-1}~M_\odot^{-1}~yr}) = 40.4^{+1.1}_{-0.3}$ (68\% CI) in this work from CXB contributions at $z\gtrsim6$ and 21-cm power spectrum limits from HERA at $z\lesssim10.35$ (hence the grey shaded region). The arrows towards high-$z$ reflect our assumption of a fixed $L_X/\text{SFR}$ relation during the early epochs. Since different works assume different SEDs (e.g. the energy range considered) which can change our constraints from CXB, a direct comparison is non-trivial but we convert the luminosity in the full X-ray band, $\SIrange{0.2}{95}{keV}$, to $\SIrange{2}{10}{keV}$ and $<\SI{2}{keV}$ for a relative comparison of the luminosity in the different energy bands.
    \label{fig:Lx_SFR_constraints}}
\end{figure}

We tabulate the full joint constraint in Table~\ref{tab:astro_param_constraints}. The case of $f_\text{X}$ is of particular interest as it is a key parameter in determining the thermal state of the IGM at $z\gtrsim6$. The joint analysis gives ${\log_{10}(f_\text{X}) = -0.10^{+1.12}_{-0.31}\left(^{+1.33}_{-2.22}\right)}$ or equivalently ${f_\text{X}=0.8^{+9.7}_{-0.4}\left(^{+16.3}_{-0.8}\right)}$ at 68\% (95\%) CI. This implies that the behaviour of high-$z$ galaxies is consistent with predictions of low metallicity HMXBs \citep[$f_\text{X}=1$;][]{2013ApJ...776L..31F} and higher than that of local star-burst galaxies \citep{2012MNRAS.426.1870M}, and we disfavour $f_\text{X}\gtrsim 17$ at 95\% CI. These constraints correspond to ${\log_{10}(L_X/\text{SFR}/\SI{}{erg~s^{-1}~M_\odot^{-1}~yr}) = 40.4^{+1.1}_{-0.3}\left(^{+1.3}_{-2.2}\right)}$ at 68\% (95\%) CI, as shown in Figure~\ref{fig:Lx_SFR_constraints}, where we place our results in the context of other works for comparison.  In particular, those using only 21-cm power spectrum upper limits have inferred higher values predicting a more heated IGM compared to our results \citep[e.g.][]{2015ApJ...809...62P,2023ApJ...945..124H,2025A&A...699A.109G}, as we shall discuss more in the next section.

\begin{table}
\centering
\renewcommand{\arraystretch}{1.3}
\caption{\label{tab:astro_param_constraints} Joint constraints from \citet{2025MNRAS.542.2292D} on the seven astrophysical parameters $\{V_c, \tilde{f}_{\star,\twoi}, M_0, \alpha_\star, \beta_\star, f_\text{X}, f_\text{r}\}$  using all the observational datasets, shown here for ease of reference. The errors are quoted at 68\% (95\%) CI around the weighted mean.}
\begin{tabular*}{\linewidth}{@{\extracolsep{\fill}}lc}
\hline
\textbf{Parameter} & \textbf{Constraint} \\
\hline
$\log_{10}(V_c/\SI{}{km/s})$ & $\lesssim 1.40\, (1.79)$ \\
$\implies V_c\,(\SI{}{km/s})$ & $\lesssim 25.2\, (62.0)$  \\
\hline
$\log_{10}(\tilde{f}_{\star,\twoi})$ & $-2.4^{+0.9}_{-0.4}\,\left(^{+1.0}_{-1.2}\right)$ \\
$\log_{10}(M_0/\text{M}_\odot)$ & $9.7^{+0.8}_{-0.6}\,\left(^{+1.1}_{-1.2}\right)$ \\
$\alpha_\star$ & $1.1^{+0.2}_{-0.2}\,\left(^{+0.4}_{-0.4}\right)$ \\
$\beta_\star$ & $2.9^{+0.9}_{-0.7}\,\left(^{+1.5}_{-1.6}\right)$ \\
\hline
$\log_{10}(f_\text{X})$ & $-0.10^{+1.12}_{-0.31}\,\left(^{+1.33}_{-2.22}\right)$ \\
$\implies \log_{10}(L_X/\text{SFR}/\SI{}{erg~s^{-1}~M_\odot^{-1}~yr})$ & $40.4^{+1.1}_{-0.3}\,\left(^{+1.3}_{-2.2}\right)$ \\
$\implies f_\text{X}$ & $0.8^{+9.7}_{-0.4}\,\left(^{+16.3}_{-0.8}\right)$ \\
\hline
$\log_{10}(f_\text{r})$ & $\lesssim 1.23\, (2.39)$ \\
$\implies f_\text{r}$ & $\lesssim 16.9\, (243.7)$ \\
\hline
\end{tabular*}
\end{table}

\section{Results} 
\label{sec:results}

The joint constraints on the astrophysical parameters described in Section~\ref{sec:astro_constraints} allow us to derive various properties of the IGM and star-formation in the early Universe. In order to visualize these functional constraints, we load the \polychord\ posterior samples (which are of order tens of thousands in number) using \anesthetic\ \citep{2019JOSS....4.1414H} and compress them to $N\sim500$ samples of equal weights\footnote{Although we compress the posterior samples to $N\sim500$ to make plotting tractable, all histograms or statistics are derived from the full posterior distribution compressed to the `maximum channel capacity' (see \anesthetic\ documentation for details).}. These samples are then used to compute the functional posteriors of the quantities of interest via emulators, trained in a similar manner to \citet{2025MNRAS.542.2292D}, as we shall describe in the following sections.

\subsection{Constraints on IGM and star-formation}
\label{sec:derived_constraints}

\begin{table*}
\centering
\caption{\label{tab:TK_TS_constraints} Constraints on the IGM kinetic temperature $T_\text{K}$, spin temperature $T_\text{S}$, and adiabatic cooling limit $T_\text{K,adiabatic}$ (top section); and on the radio background temperature $T_\text{rad}(\nu_{21})$, $T_\text{S}/T_\text{rad}(\nu_{21})$, and CMB temperature $T_\text{CMB}$ (bottom section) at $z=6,8,10,12.5,15$. For each quantity, we show the prior range and the posterior range, both at 95\% confidence interval (CI). The numbers in bold for the $T_\text{S}/T_\text{rad}$ column highlight the optimistic predictions for the observability of the 21-cm signal.}
\begin{tabular*}{1\linewidth}{@{\extracolsep{\fill}}cccccc}
\hline
& \multicolumn{2}{c}{$T_\text{K}$ (K)} & \multicolumn{2}{c}{$T_\text{S}$ (K)} & $T_\text{K,adiabatic}$ (K) \\
Redshift & Prior & 95\% CI & Prior & 95\% CI & Value \\
\hline
$z=6$    & $[2.5, 165043.8]$ & $[19.8, 2077.9]$ & $[2.7, 40110.9]$ & $[19.0, 1258.1]$ & $1.0$ \\
$z=8$    & $[1.7, 106530.4]$ & $[3.7, 349.5]$ & $[2.7, 25677.9]$ & $[4.3, 359.2]$ & $1.6$ \\
$z=10$   & $[2.4, 49516.9]$ & $[2.5, 66.2]$ & $[3.2, 11624.1]$ & $[3.1, 73.3]$ & $2.4$ \\
$z=12.5$ & $[3.6, 6956.2]$ & $[3.6, 16.0]$ & $[4.4, 2939.6]$ & $[4.5, 19.0]$ & $3.6$ \\
$z=15$   & $[5.0, 704.8]$ & $[5.0, 7.7]$ & $[5.6, 578.3]$ & $[6.4, 33.9]$ & $5.0$ \\
\hline
& \multicolumn{2}{c}{$T_\text{rad}(\nu_{21})$ (K)} & \multicolumn{2}{c}{$T_\text{S}/T_\text{rad}(\nu_{21})$} & $T_\text{CMB}$ (K) \\
Redshift & Prior & 95\% CI & Prior & 95\% CI & Value \\
\hline
$z=6$    & $\leq 93900.5$ & $\leq 101.0$ & $[0.0, 592.7]$ & $[0.8, 54.7]$ & $19.1$ \\
$z=8$    & $\leq 43163.3$ & $\leq 66.5$ & $[0.0, 222.0]$ & $[0.2, 11.9]$ & $24.5$ \\
$z=10$   & $\leq 18636.8$ & $\leq 50.5$ & $[0.0, 172.0]$ & $[0.1, 2.2]$ & $30.0$ \\
$z=12.5$ & $\leq 5891.6$ & $\leq 44.7$ & $[0.0, 55.0]$ & $\mathbf{[0.1, 0.5]}$ & $36.8$ \\
$z=15$   & $\leq 1652.7$ & $\leq 46.8$ & $[0.0, 11.1]$ & $\mathbf{[0.1, 0.8]}$ & $43.6$ \\
\hline
\end{tabular*}
\end{table*}

\begin{figure*}
   \includegraphics[width=1\linewidth]{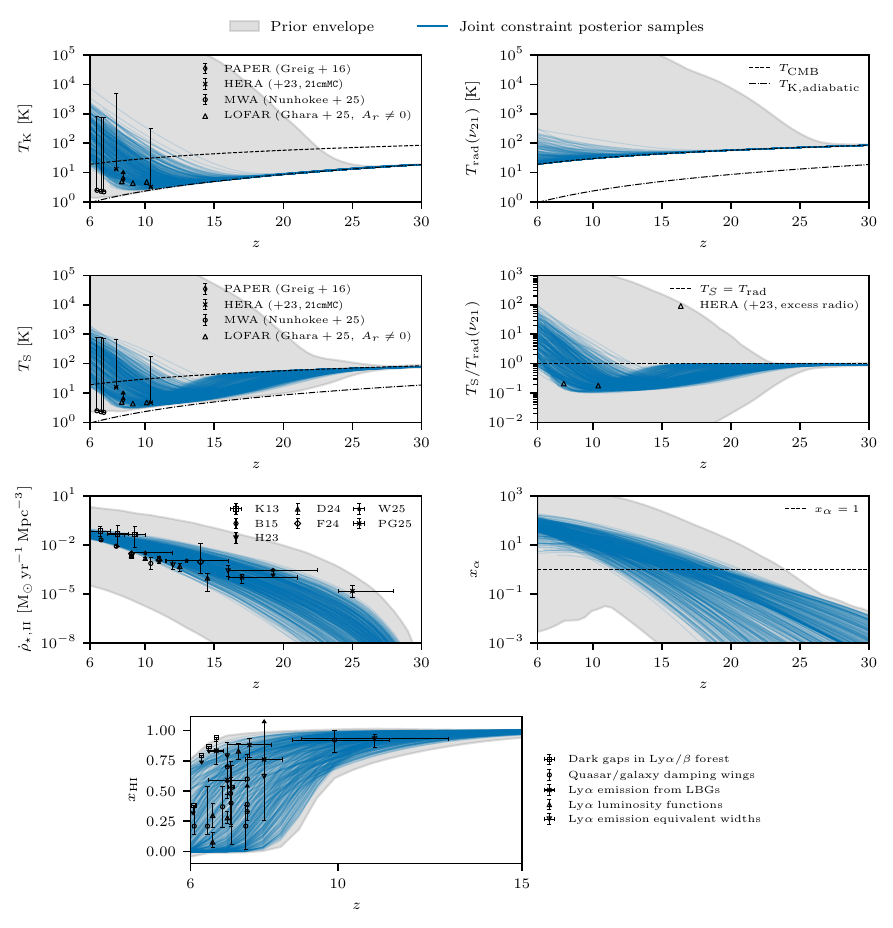}
    \caption{Evolution of various quantities related to the IGM and star-formation in the redshift range $z\approx6-30$. The panels from top-left to bottom-right show the kinetic temperature of neutral IGM ($T_\text{K}$), the radio background temperature at the rest-frame 21-cm wavelength ($T_\text{rad}$), the spin temperature ($T_\text{S}$), the ratio of spin to radio background temperature ($T_\text{S}/T_\text{rad}$), the Pop~II star-formation rate density (SFRD, $\dot{\rho}_{\star,\twoi}$), and the neutral IGM fraction ($x_\text{HI}$) at the bottom. The grey regions show the priors in this functional space, while the solid lines are $N\sim500$ samples from the joint posterior distribution. In the top panels, we also show the latest constraints on neutral IGM $T_\text{K}$ from PAPER, \citep{2015ApJ...809...62P,2016MNRAS.455.4295G}, HERA+23 \citep[][]{2023ApJ...945..124H}, LOFAR \citep[][their $A_r\neq 0$ model]{2025A&A...699A.109G}, MWA \citep{2025ApJ...989...57N}. In case of the HERA+23 constraints, we plot the results from their \texttt{21cmMC} model \citep[generated using \textsc{21cmFAST};][]{2011MNRAS.411..955M} for $T_\text{K}$ and excess radio model  (generated using \simcode) for $T_\text{S}/T_\text{rad}$. We also plot the same constraints for $T_\text{S}$, assuming $T_\text{S}=T_\text{K}$ for MWA (although not stated explicitly in their work, it is reasonable to assume saturated WF coupling at $z\lesssim 7$). In the panel for the SFRD $\dot{\rho}_{\star,\twoi}$, we show various inferences from literature using HST/JWST UVLFs \citep[][]{2015ApJ...803...34B,2023ApJS..265....5H,2024MNRAS.533.3222D,2024ApJ...969L...2F,2025ApJ...992...63W,2025ApJ...991..179P}, and gamma-ray bursts \citep{2013arXiv1305.1630K}. Finally, in the bottom panel for neutral fraction $x_\text{HI}$, we show constraints from dark gaps in Ly$\alpha/\beta$ forest \citep{2015MNRAS.447..499M,2023ApJ...942...59J}, from Ly$\alpha$ damping wings \citep{2017MNRAS.466.4239G,2019MNRAS.484.5094G,2018ApJ...864..142D,2018Natur.553..473B,2020ApJ...896...23W,2020ApJ...897L..14Y,2024ApJ...969..162D,2024ApJ...971..124U}, fraction of Lyman-break galaxies showing Ly$\alpha$ emission \citep{2018ApJ...856....2M,2019MNRAS.485.3947M,2019ApJ...878...12H,2022MNRAS.517.3263B}, from Ly$\alpha$ luminosity functions \citep{2021ApJ...919..120M,2022ApJ...926..230N}, and from Ly$\alpha$ equivalent width distributions \citep{2024ApJ...967...28N}. In this work, we do not constrain $x_\text{HI}$ anymore than the flat $3\sigma$ prior on $\tau_\text{CMB}$ from $\textit{Planck 2018}$, hence the overlap of grey shaded region and blue lines, but we show a variety of constraints from literature for completeness and to highlight the uncertainty in EoR evolution.
    \label{fig:derived_constraints}}
\end{figure*}

We first discuss the constraints on the kinetic temperature of the neutral IGM ($T_\text{K}$), the radio background temperature ($T_\text{rad}$), the 21-cm spin temperature ($T_\text{S}$), the Pop~II star-formation rate density (SFRD; $\dot{\rho}_{\star,\twoi}$) and the neutral IGM fraction ($x_\text{HI}$) in the redshift range $z=\SIrange{6}{30}{}$. All quantities are trained on \simcode\ outputs using the \textsc{TensorFlow} \citep{2016arXiv160304467A} based package \globalemu\ \citep{2021MNRAS.508.2923B}, with a test/train split of 1/9, a learning rate of $10^{-3}$, and the $\tanh$ activation function (except for the $\dot{\rho}_{\star,\twoi}$ emulator which has a $\text{ReLU}$ activation function). The $T_\text{K}$, $T_\text{S}$, and $T_\text{rad}$ emulators have an architecture of 4 hidden layers with 32 nodes each, $\dot{\rho}_{\star,\twoi}$ emulator has 4 hidden layers with 16 nodes each, and the $x_\alpha$ and $x_\text{HI}$ emulators have 3 hidden layers with 32 nodes each, all yielding an error of $\lesssim 10\%$ at 65th percentile of the test data in log-space.

We show the posteriors from the joint constraints for each of the above quantities in Figure~\ref{fig:derived_constraints}, and describe them in detail below:

\begin{figure*}
   \includegraphics[width=1\linewidth]{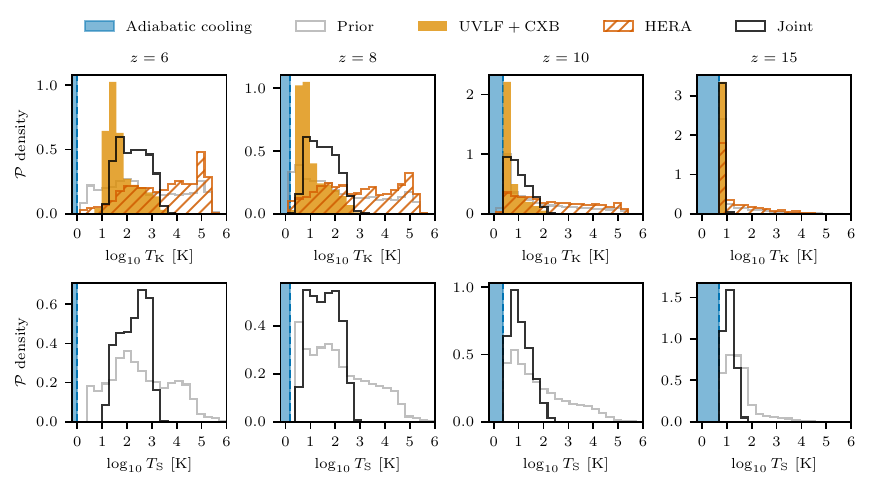}
    \caption{PDFs of the IGM kinetic temperature $T_\text{K}$ (top row) and spin temperature $T_\text{S}$ (bottom row) at $z=6,8,10$ and $z=15$ in the joint analysis. Note that the histograms are not from one simulation, but denote many realizations of the Universe with different astrophysical parameters sampled from the posterior. The blue region denotes the adiabatic cooling limit, while the grey and black histograms are the prior and joint posteriors from our models and analysis respectively. The constraints on $T_\text{K}$ are driven by a combination of UVLF + CXB data disfavouring high $f_\text{X}$ values (shown in solid yellow histogram), and HERA disfavouring low $f_\text{X}$ values (shown in hatched orange histograms). It can be seen that we only get marginal heating at $z\gtrsim 15$, while it is guaranteed at $z\lesssim 8$.
    \label{fig:TK_TS_distributions}}
\end{figure*}

\begin{itemize}
\item \textbf{Kinetic temperature of neutral IGM}, $T_\text{K}$ (first row, left panel of Figure~\ref{fig:derived_constraints}): As the main source of IGM heating in our models is via X-rays from low metallicity HMXBs (although we do include Ly$\alpha$ heating which can be important for weak X-ray efficiencies), the constraints on $T_\text{K}$ are dictated by the constraint on $f_\text{X}$. By tightly constraining the X-ray heating efficiency of galaxies from both ends of the prior, the joint analysis favours a late heating scenario in which the neutral IGM remains largely adiabatically cooled\footnote{For reference, the adiabatic cooling limit as plotted/mentioned throughout this work is $T_\text{K,adiabatic}(z)\approx \SI{50.67}{K}\times(1+z)^2/(1+50)^2$.} at $z = 15$ (${\SI{5.0}{K} \lesssim T_\text{K} \lesssim \SI{7.7}{K}}$), reaches modest temperatures by $z=10$ (${\SI{2.5}{K} \lesssim T_\text{K} \lesssim \SI{66}{K}}$), and is assuredly heated above the adiabatic limit at $z=8$ (${\SI{3.7}{K} \lesssim T_\text{K} \lesssim \SI{350}{K}}$) and $z=6$ (${\SI{20}{K} \lesssim T_\text{K} \lesssim \SI{2078}{K}}$) at 95\% CI. Indeed, we do not rule out a weakly heated IGM even at low redshifts (the so-called `cold reionization scenario') on account of high $f_\text{X}$ being disfavoured by UVLF + CXB data, except towards the end of the EoR at $z=6$ when $T_\text{K} \gtrsim T_\text{CMB}=\SI{19}{K}$. These results, tabulated in Table~\ref{tab:TK_TS_constraints}, are in good agreement with the IGM evolution inferred in \citet[][]{2017ApJ...840...39M} who use an empirical model calibrated using SFRDs and the CXB (although we allow for a wider range of values; e.g. they infer $\SI{300}{K}\lesssim T_\text{K}(z=6) \lesssim \SI{600}{K}$). Our results also generally agree with the inference from PAPER \citep[][$T_\text{K} \gtrsim \SI{6}{K}$ at $z=8.4$, although at 60\% CI]{2015ApJ...809...62P,2016MNRAS.455.4295G}, and LOFAR \citep[][$T_\text{K} \gtrsim \SI{4.4}{K}$ at $z=9.1$]{2025A&A...699A.109G}, but are less consistent with HERA+23 \citep[][$T_\text{K} \gtrsim \SI{13}{K}$ at $z=7.9$ at 95\% CI]{2023ApJ...945..124H}. It is worth noting that the HERA+23 analysis using \texttt{21cmMC} does not include Ly$\alpha$ heating, which is included in our analysis. Since the aforementioned works use 21-cm power spectrum upper limits without including CXB data, they infer higher $f_\text{X}$ values and thus allow for a more heated IGM. Similarly, previous analyses using \simcode\ \citep{2024MNRAS.531.1113P,2024MNRAS.529..519G} that include both HERA and CXB data, do not include UVLFs and thus have weaker upper limits on $f_\text{X}$ and consequently $T_\text{K}$. The joint constraints here improve upon them, providing a more complete picture of the heated IGM during the EoR. Furthermore, at $z\lesssim7$, the recent limits and analysis from MWA \citep{2025ApJ...989...57N} provide evidence for a heated IGM at those epochs, but since they are less constraining than HERA, we find comfortable agreement with their results.

To illustrate the results further, we plot the posterior distribution of $T_\text{K}$ at $z=6,8,10$ and $z=15$ in Figure~\ref{fig:TK_TS_distributions} (top row), clearly showing the contribution of the CXB data in constraining high $T_\text{K}$\footnote{Note that the since CXB and UVLF data are combined in the analysis here, there is \textit{some} heating even with the lowest $f_\text{X}$ due to guaranteed star-formation. It is only when there is very low SFE (or equivalently SFR, which UVLFs disfavour) combined with weak heating that the adiabatic prior limit is recovered.}.

\item \textbf{Radio background temperature at rest-frame 21-cm wavelength}, $T_\text{rad}(\nu_{21})$ (first row, right panel of Figure~\ref{fig:derived_constraints}): The radio background temperature, as sourced by the CMB and radio galaxies in our model, is determined by the emission efficiency $f_\text{r}$. The joint analysis favours $f_\text{r} \lesssim 16.9\,(243.7)$ at 68\% (95\%) CI, with contributions from UVLF + CRB and HERA data. Hence, we infer upper limits of $T_\text{rad} \lesssim \SI{47}{K}, \SI{51}{K}, \SI{67}{K}$ and $\SI{101}{K}$ at $z=15, 10, 8$ and $z=6$ respectively at 95\% CI, as shown in Table~\ref{tab:TK_TS_constraints}. 

\item \textbf{21-cm spin temperature}, $T_\text{S}$ (second row of Figure~\ref{fig:derived_constraints}): Once star-forming galaxies become abundant in the Universe (as informed by HST and JWST), the 21-cm spin temperature $T_\text{S}$ couples to the matter temperature $T_\text{K}$ via Ly$\alpha$ radiation from galaxies (see discussion on $x_\alpha$ constraints later). The constraints from the joint analysis are tabulated in Table~\ref{tab:TK_TS_constraints} and also shown in the second row of Figure~\ref{fig:TK_TS_distributions}. The conclusions are similar to that of $T_\text{K}$, and although the $T_\text{S}$ constraints are useful in their own regard, we observe the 21-cm signal in contrast to the radio background. Thus the most informative quantity is the ratio of the spin temperature to the radio background temperature $T_\text{S}/T_\text{rad}$, as shown in the second row, right panel of Figure~\ref{fig:derived_constraints}. The joint analysis gives us a striking result. With efficient coupling and a weakly heated IGM at $z\lesssim15$, there is definitive contrast between $T_S$ and $T_\text{rad}$ as evidenced by the $T_\text{S}/T_\text{rad} \lesssim 1$ at $z \approx 12.5$. This has important implications for the 21-cm signal and its detectability, and we discuss this in detail in Section~\ref{sec:21cm_signal_constraints}. At lower redshifts, $z\lesssim10.5$, we infer a similar $T_\text{S}/T_\text{rad}$ to HERA+23 \citep[][although we allow for lower values on account of weaker heating]{2023ApJ...945..124H}.

\item \textbf{Pop~II SFRD}, $\dot{\rho}_{\star,\twoi}$ (third row, left panel of Figure~\ref{fig:derived_constraints}): As described in Section~\ref{sec:astro_constraints}, the Pop~II SFE is constrained by the UVLF data \citep[see Figure~7 in][for the functional posteriors]{2025MNRAS.542.2292D}, and shows preference for redshift independence at $z\lesssim10$ and a strong redshift evolution at $z\gtrsim10$. The SFRD evolution inferred from the joint analysis, which is a compromise between the two redshift regimes, sits above the HST and JWST observations of Lyman-break galaxies $z\lesssim15$ \citep[e.g.][]{2015ApJ...803...34B,2024MNRAS.533.3222D}. This is expected as the SFRD shown in Figure~\ref{fig:derived_constraints} is the total SFRD of Pop~II star-forming galaxies, including DM halos down to the minimum circular velocity, $V_c$. This can be as low as $M_h\sim 10^6 M_\odot$ at $z=15$ and $M_h\sim 10^5 M_\odot$ at $z=30$ for molecular cooling halos, well outside the faintest observable magnitudes. As an example, assuming a SFE of 100\%, this would correspond to magnitudes of $M_\text{UV}\sim -11.5$ and $M_\text{UV}\sim -10$ respectively in the absence of lensing magnification. A more direct comparison to HST/JWST observations would be to infer the SFRD by integrating the UVLF at $M_\text{UV} < -17$ \citep[as in Figure~9 of][]{2025MNRAS.542.2292D}. SFRD inferred from gamma-ray bursts \citep{2013arXiv1305.1630K}, which are sensitive to obscured star-formation unlike HST/JWST observations, are also consistent with our results in Figure~\ref{fig:derived_constraints}. Recent estimates suggest that dust-obscured star-formation can account for as high as $\sim 40\%$ of all star-formation at $z\sim 6$ \citep{2025arXiv250805740B}.

\item \textbf{Ly$\alpha$ coupling}, $x_\alpha$ (third row, right panel of Figure~\ref{fig:derived_constraints}): The Ly$\alpha$ flux from Pop~II star-forming galaxies facilitates the WF coupling of the spin temperature $T_\text{S}$ to the kinetic temperature of the IGM $T_\text{K}$, quantified by the coupling coefficient as defined in Equation~\ref{eqn:TS_equation}, and becomes saturated when $x_\alpha \gg 1$. We derive constraints of $x_\alpha \gtrsim 0.1, 0.8, 4.8, 11.2$ and $13.3$ at $z=15, 12.5, 10, 8$ and $z=6$ respectively at 95\% CI.

\item \textbf{Neutral IGM fraction}, $x_\text{HI}$ (bottom panel of Figure~\ref{fig:derived_constraints}): We do not constrain the ionization of the IGM in our models anymore than the prior, which is set to be uniformly sampled from $3\sigma$ around the \textit{Planck 2018} optical depth $\tau_\text{CMB}\in[0.033, 0.075]$. Of all datasets, the only one with the potential to constrain the neutral fraction is the 21-cm power spectrum limits from HERA. HERA marginally prefers a higher $\tau_\text{CMB}$ (i.e., early reionization at $z\approx 8$) as an alternative solution to the vanishing 21-cm power spectrum in extremely weak heating cases ($f_\text{X}\lesssim 0.1$, which is why the $f_\text{X}$ PDF does not fall to zero at the low end of the prior in Figure~\ref{fig:param_constraints}). Beyond this degeneracy, HERA is no more informative than the \textit{Planck 2018} prior on $x_\text{HI}$, hence, we retrieve a prior-dominated posterior. We comment on the prospects for future work in this realm in Section~\ref{sec:reionization}.
\end{itemize}

\begin{figure*}
   \includegraphics[width=1\linewidth]{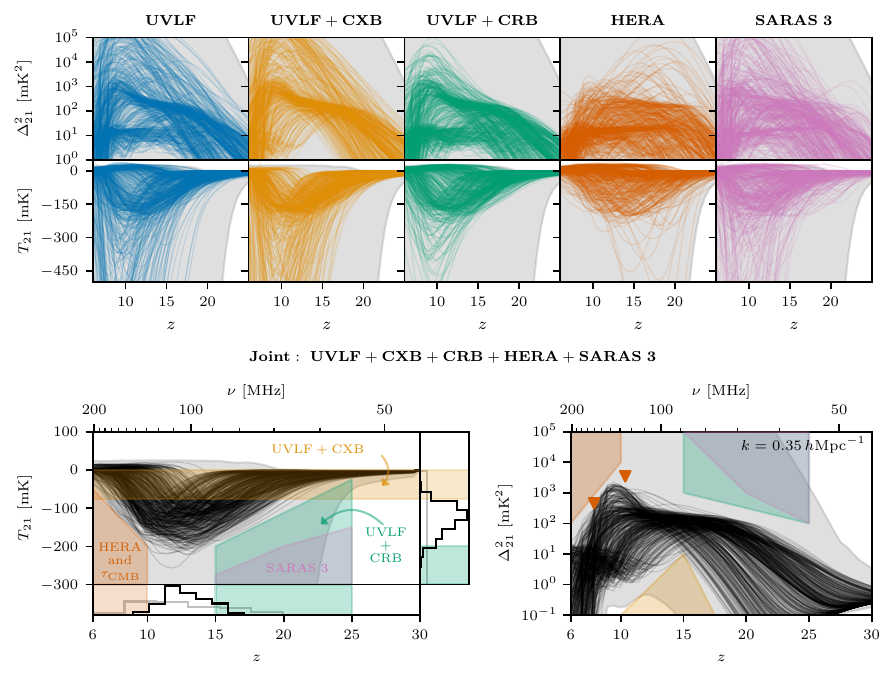}
    \caption{Constraints on the 21-cm global signal $T_{21}$ and power spectrum $\Delta_{21}^2$ from the individual datasets (top row) and the joint analysis (bottom row). The grey regions show the prior space, while the solid lines are $N\sim500$ samples from the respective posterior distributions. In the joint constraint panels, we highlight qualitatively regions of the prior space (as the filled coloured regions) that are disfavoured by the different datasets. The global signal panel also show histograms of the redshift/timing (along $x$-axis) and depth (along $y$-axis) of the absorption trough minima, while the power spectrum panel shows the 21-cm power spectrum limits from HERA \citep[][in orange]{2023ApJ...945..124H}. Most notably, via the constraints on heating efficiency $f_\text{X}$ and IGM temperature $T_\text{K}$ from the UVLF + CXB data, we find a non-vanishing global signal and power spectrum at $z\sim 10-15$ (see Section~\ref{sec:21cm_signal_constraints} for quantified constraints). For the special cases of joint analysis where we exclude certain datasets, see Figure~\ref{fig:21cm_constraints_joint}.
    \label{fig:21cm_constraints}}
\end{figure*}

\subsection{Constraints on 21-cm global signal and power spectrum}
\label{sec:21cm_signal_constraints}

\begin{figure*}
   \includegraphics[width=1\linewidth]{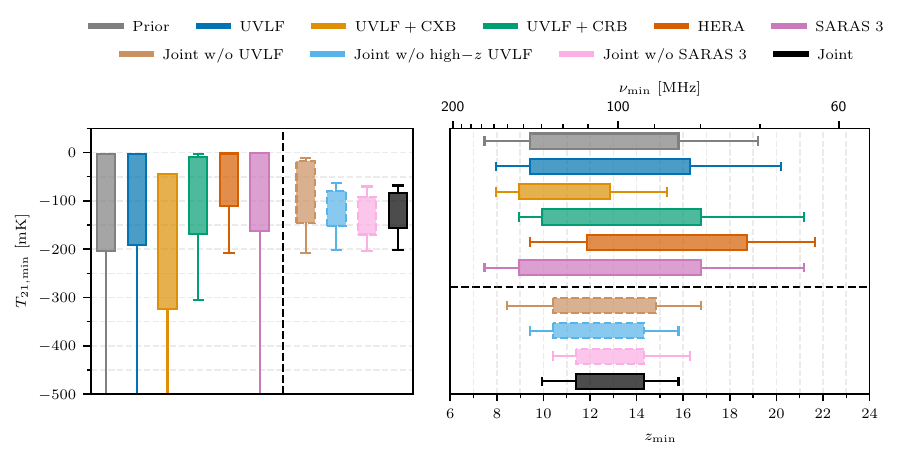}
    \caption{Another way to quantify the global signal constraints from individual datasets and joint analysis shown in Figure~\ref{fig:21cm_constraints} and Figure~\ref{fig:21cm_constraints_joint}. The plot shows the depth (signal amplitude; left panel) and timing (redshift; middle panel) of the 21-cm global signal absorption trough, with the dashed black line separating the joint cases. The box-plots here show the 68\% CI (boxes) and 95\% CI (whiskers) of the prior and posterior distributions.
    \label{fig:21cm_troughs}}
\end{figure*}

The constraints on the IGM temperature $T_\text{K}$ and the radio background temperature $T_\text{rad}$ are crucial for understanding the 21-cm signal. In order to quantify the 21-cm signal constraints, we improve upon the emulators for the 21-cm global signal and power spectrum from \citet{2025MNRAS.542.2292D} by re-training them with updated architectures. The former is re-trained with an architecture of 4 hidden layers with 32 nodes each (instead of 16 nodes each as used in the earlier work), still \globalemu\ with a test/train split of 1/9. The new emulator has a root-mean squared error of $\SI{9.1}{mK}$ at 68th percentile of the test data\footnote{For test data signals that have absorption amplitude of $T_\text{21,min}$ smaller than $\SI{-50}{mK}$, the RMSE is $\SI{5}{mK}$ at 68th percentile. This is consistent with a $\sim 10\%$ emulator error, which is smaller than the assumed semi-numerical errors of $\sim 20\%$.}. The latter is re-trained with the same architecture as the previous work and with \textsc{scikit-learn}'s \texttt{MLPRegressor}, but with an increased redshift range of $z=6-30$ (instead of $z=7-11$ used earlier), yielding an error of $\lesssim 20\%$ for all $k$ and $z$ bins.

Figure~\ref{fig:21cm_constraints} shows the constraints on the 21-cm global signal and power spectrum at $z=6-30$ for each dataset in the top row, and the joint analysis in the bottom row. In particular, the figure highlights the regions of the function space that are constrained by each of the datasets used in this work.

The UVLF data on its own only rules out very early, deep global signals at $z\gtrsim15$ (arising from very high SFE) and very low amplitude power spectra at $z\approx10-15$ (arising from very low SFE). These extreme cases are disfavoured by anchoring star-formation to the range allowed by the galaxy abundances (i.e. UVLF obserations). Building on this, the UVLF + CXB data combination provides a strong constraint on how efficient the heating can be before it is in conflict with the diffuse X-ray background limits. This rules out very shallow global absorption signals and suppressed power spectra at $z\sim10-15$. The UVLF + CRB combination, on the other hand, rules out high $f_\text{r}$ and thus deep signals due to an excess radio background sourced by efficient radio galaxies, which has implications for deep signals at $z\gtrsim15$. HERA, in contrast to UVLF + CXB, prefers a heated IGM to suppress high power spectra (and consequently global signal amplitude) at $z\lesssim11$. SARAS~3 prefers to push deep global signals outside of the redshift range $z\approx15-25$ (and thus disfavours high power spectra at early times). One way to achieve this is to disfavour efficient star-formation in molecular cooling halos that cause an early onset of the absorption signal (i.e., low $V_c$)\footnote{Deep signals resulting from excess radio background are mildly disfavoured by SARAS~3 as shown in the marginalized PDF for $f_\text{r}$ in Figure~\ref{fig:param_constraints}. We expect that this constraint would be strengthened if SARAS~3 data were combined with UVLF data, so that degenerate low SFE -- high $f_\text{r}$ combinations are ruled out.}.

\begin{figure}
   \includegraphics[width=1\linewidth]{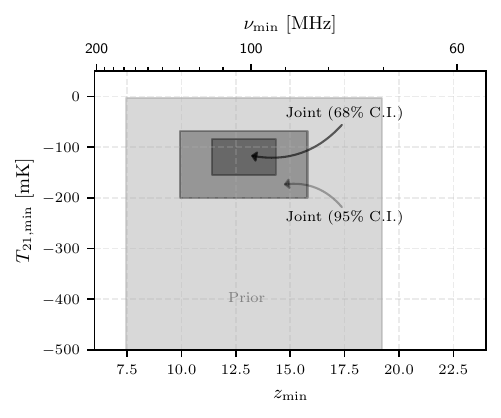}
    \caption{A `treasure map' of the discovery space for the 21-cm global signal absorption trough with the priors in light grey (at 95\% CI), and the joint posterior (at 68\% and 95\% CI, as annotated) in darker grey. The exact values are shown in Table~\ref{tab:21cm_constraints}.
    \label{fig:21cm_troughs_map}}
\end{figure}

The case of joint analysis is most interesting because it provides stringent (however, model-dependent) constraints on the 21-cm signal. In the global signal panel of Figure~\ref{fig:21cm_constraints}, the histograms along the $x$ and $y$ axes show the timing and depth of the absorption trough respectively. The constraints predict the global signal absorption trough to lie at $9.9\lesssim z_\text{min}\lesssim 15.8$ (or equivalently $\SI{130}{MHz}\lesssim\nu_\text{min}\lesssim\SI{85}{MHz}$) with a depth of:
\begin{align*}
\SI{-156}{mK} &\lesssim T_{21,\mathrm{min}}\lesssim\SI{-84}{mK} \text{ at 68\% CI}, \\
\SI{-201}{mK} &\lesssim T_{21,\mathrm{min}}\lesssim\SI{-68}{mK} \text{ at 95\% CI},
\end{align*}
and $T_\text{21,min}<\SI{-278}{mK}$ and $T_\text{21,min}>\SI{-33}{mK}$ are ruled out at 99.99\% CI. We also show these constraints in the form of box-plots in Figure~\ref{fig:21cm_troughs}, for all individual and joint datasets, and in the form of a `discovery space' map ($T_\text{21,min}$ vs. $z_\text{min}$) in Figure~\ref{fig:21cm_troughs_map} for the joint analysis.

The power spectrum panel of Figure~\ref{fig:21cm_constraints} shows $\Delta_{21}^2(z)$ in the $k$-bin where HERA limits are strongest ($k \approx 0.35\,h\text{Mpc}^{-1}$). Most interestingly, we infer a non-vanishing power spectrum at $z=15$ of:
\begin{align*}
\SI{50}{mK^2} &\lesssim \Delta_{21}^2(z=15)\lesssim\SI{157}{mK^2} \text{ at 68\% CI}, \\
\SI{8.7}{mK^2} &\lesssim \Delta_{21}^2(z=15)\lesssim\SI{197}{mK^2} \text{ at 95\% CI}.
\end{align*}
We also constrain the power spectrum at $z=8, 10, 20, 25$ as tabulated in Table~\ref{tab:21cm_constraints}. The power spectrum at $z=8$ and $z=10$ is also non-vanishing, but the constraints are weaker, while the power spectrum $z=20$ and $z=25$ is consistent with zero, but with informative upper limits. The joint analysis thus provides a lower limit on the 21-cm global signal depth and a non-vanishing power spectrum that is not prior-limited. This is a significant step forward in our understanding of the expected range of 21-cm signals.

Thus, we have constrained the 21-cm signal within our model parameterization, and, for the first time, inferred a lower limit on the global signal depth and non-vanishing power spectrum that is not prior-limited. These constraints would not be possible without the inclusion of the UVLF data, as shown in brown in Figure~\ref{fig:21cm_troughs} and Figure~\ref{fig:21cm_constraints_joint} (top-row), and indeed provide an optimistic view of the 21-cm signal at high-$z$. Before we discuss the caveats of this inference, we test the robustness of these constraints by performing the joint analysis without the high-$z$ UVLF data from JWST, and without the SARAS~3 data which relies on proper foreground subtraction.

\subsubsection{In the absence of high-$z$ UVLFs}
\label{sec:without_highz_UVLFs}

In light of the abundance of bright galaxies seen by JWST at $z\gtrsim10$, there is increasing interest in the impact of how different galaxy formation models affect our understanding of the 21-cm signal \citep[e.g.][]{2023MNRAS.525..620C,2024MNRAS.532..149L,2025MNRAS.543.1058A}. The results presented in this work do not change much in the absence of high-$z$ UVLFs ($z\gtrsim 10$), as shown in light-blue in Figure~\ref{fig:21cm_troughs} and Figure~\ref{fig:21cm_constraints_joint}. This is expected even though the constraint on $\beta_\star$ changes, meaning the SFE evolves with redshift. The SFRD and UVLFs are still sensitive to $\beta_\star$, but the constraints on $f_\text{X}$ and $f_\text{r}$ from the diffuse X-ray and radio backgrounds respectively largely come from galaxies at $z\sim6$, where $\beta_\star$ has no effect (see Footnote~\ref{foot:fXfr_robustness}). There is a marginal shift in the median global signal trough by $\Delta z\sim0.5$ to lower $z$, which is consistent with the analysis of \citet{2024MNRAS.532..149L} in their `feedback-free burst' model for galaxies (which is similar to our redshift-dependent enhancement of the SFE). This is because a fixed SFE that is not enhanced in the early Universe means a less efficient WF coupling at high-$z$, and thus a later onset of the absorption signal.

\begin{table*}
\centering
\renewcommand{\arraystretch}{1.3}
\caption{\label{tab:21cm_constraints} Constraints on the 21-cm global signal $T_{21}$ (minimum absorption) and power spectrum $\Delta_{21}^2$ at $k=0.35\,h\,\mathrm{Mpc}^{-1}$ at selected redshifts. For each quantity, we show the prior range (at 95\% CI) and the joint posterior range (at 68\% and 95\% CI). The numbers in bold highlight the optimistic predictions for the observability of the 21-cm signal.}
\begin{tabular*}{\linewidth}{@{\extracolsep{\fill}}cccc}
\hline
\multicolumn{4}{c}{\textbf{21-cm global signal} (in mK)} \\
Quantity & Prior & 68\% CI & 95\% CI \\
\hline
$T_{21,\mathrm{min}}$ & [-3796, -2] mK & \textbf{[-156, -84]} mK & \textbf{[-201, -68]} mK \\
$z_\text{min}$ & [7.5, 19.2] & [11.4, 14.3] & [9.9, 15.8] \\
$\nu_\text{min}$ & [168, 70] MHz & [115, 93] MHz & [130, 85] MHz \\
\hline
\multicolumn{4}{c}{\textbf{21-cm power spectrum} (in mK$^2$)} \\
Quantity & Prior & 68\% CI & 95\% CI \\
\hline
$\Delta_{21}^2(z=8)$ & $[0, 4.15 \times 10^5]$ & $[0.004, 16.4]$ & $[0.004, 180]$ \\
$\Delta_{21}^2(z=10)$ & $[0.18, 1.21 \times 10^6]$ & $[0.47, 179]$ & $[0.47, 1020]$ \\
$\Delta_{21}^2(z=15)$ & $[0.06, 1.86 \times 10^5]$ & $\mathbf{[50, 157]}$ & $\mathbf{[8.7, 197]}$ \\
$\Delta_{21}^2(z=20)$ & $[0.005, 3070]$ & $[0.01, 35]$ & $[0.01, 80]$ \\
$\Delta_{21}^2(z=25)$ & $[0.051, 19.3]$ & $[0.056, 0.90]$ & $[0.056, 6.2]$ \\
\hline
\end{tabular*}
\end{table*}

\subsubsection{In the absence of SARAS~3}
\label{sec:without_SARAS3}

As described in Section~\ref{sec:observational_datasets}, our Bayesian likelihood for the SARAS~3 data employs a sixth-order log-log polynomial to model the Galactic and extragalactic foreground. Evidently, a small shift in any of the polynomial coefficients can lead to large (tens of mK level) changes in the global signal posteriors that can bias our astrophysical parameter constraints. This was demonstrated in \citet[][Figure C3]{2025MNRAS.542.2292D}, where the leading polynomial coefficients are degenerate with the constraint on $V_c$. Thus, we perform a joint analysis without the SARAS~3 data to ensure a wrong choice of foreground model is not affecting our results, shown in light-pink in Figure~\ref{fig:21cm_troughs} and Figure~\ref{fig:21cm_constraints_joint}. The constraint on the global signal depth remains unaffected, but the timing shifts by $\Delta z\sim0.5$ to higher $z$. This is because the weak constraint on models with low $V_c$, disfavouring molecular cooling halos, disappears (see first panel of Figure~\ref{fig:param_constraints_joint}) and star-formation can begin earlier, leading to an earlier onset of the absorption signal.

\subsection{Model dependency and caveats}
\label{sec:model_dependency}

The results presented in this work offer promising constraints on the 21-cm signal and optimistic prospects for its detection. \textit{However, they do not constitute an indirect detection of the 21-cm signal}. They are model-dependent inferences based on our astrophysical parameterization and come with some assumptions and caveats. In particular, we highlight how our results may change with different assumptions about the X-ray emissions and inclusion of a flexible Pop~III star-formation model.

\subsubsection{X-ray emission at high-$z$}

Our knowledge of high-$z$ galaxies and their X-ray emission properties is still limited. Thus the lower limit on the 21-cm signal (driven by UVLF + CXB data) is dependent on the X-ray model. The assumptions we make about the X-ray emissions in this work include:

\begin{enumerate}
\item \textit{X-ray emissions are dominated by HMXBs.} Due to their short lifetimes, formation in high (or bursty) star-forming and low metallicity environments, they are expected to dominate the X-ray emissions at $z\gtrsim6$ \citep{2013ApJ...776L..31F,2016ApJ...825....7L,2017ApJ...840...39M}. With the recent observations of JWST showing an increasing population of AGN at earlier times \citep[e.g.][albeit many without X-ray detections]{2023MNRAS.525.1353J,2023ApJ...959...39H,2024NatAs...8..126B,2024A&A...691A.145M,2024ApJ...977..250F}, the contribution of AGN to the X-ray background at high-$z$ remains uncertain \citep[in particular, understanding the X-ray quiet `Little Red Dots' discovered by JWST; e.g.][]{2025ApJ...979..138H}. Furthermore, other sources of heating, such as cosmic rays produced by Pop~II and Pop~III supernovae may be able to heat the IGM above the CMB temperature by $z\sim10$ already \citep{2015MNRAS.454.3464S,2017MNRAS.469..416L,2019MNRAS.483.5329J,2023MNRAS.526.4262G}, which we do not explore in this work.
\item \textit{X-ray SED is fixed across cosmic time}. We assume that the X-ray SED of HMXBs is fixed to \citet[][their mean model at $z=15.34$]{2013ApJ...776L..31F} in the energy range $\SIrange{0.2}{95}{keV}$, which is a relatively hard spectrum. Since the X-ray SED slope and minimum/maximum energy cutoffs remain essentially unconstrained so far \citep[e.g.][]{2024MNRAS.531.1113P,2025JCAP...10..047K}, a softer spectrum \citep[from hot ISM instead of HMXBs, e.g.][]{2014MNRAS.443..678P} may be able to provide more heating efficiency whilst avoiding exceeding the CXB limits. This has been explored previously in \citet{2017MNRAS.464.3498F}, including both soft and hard X-ray SEDs and contributions from high-$z$ quasars, but using only CXB data as opposed to the multi-wavelength data used in this work.
\item \textit{$L_X/\text{SFR}$ is constant over cosmic time}, scaled by the free parameter $f_\text{X}$ \citep[normalized to the theoretical prediction for low metallicity HMXBs from][]{2013ApJ...776L..31F} that we constrain. Assuming this star-formation scaling, the X-ray emissions integrated across the simulation volume and time (accounting for redshifting spectra) contribute to the diffuse CXB at $z=0$. The largest contributions come from galaxies at the lowest redshift ($z\sim 6$) in our simulations where the SFRD is largest (anchored by the UVLF data) and spectra are redshifted the least. Hence, the $f_\text{X}$ constraint from CXB is most robust for populations close to $z\sim6$. The constraints become increasingly weaker on the higher-$z$ populations, where the spectra are redshifted and star-forming galaxies contribute less to the CXB. In particular, we expect our 21-cm signal constraints to be affected if the $L_X/\text{SFR}$ relation strongly depends on the metallicity as stellar populations evolve from primordial metallicities of $Z/Z_\odot < 10^{-6}$ to $\sim10^{-2}$ over cosmic time. This regime lies outside of observed ranges, but there is growing evidence of a strong metallicity dependence \citep[e.g.][]{2015A&A...579A..44D,2016MNRAS.457.4081B,2019ApJ...885...65F,2024ApJ...977..189L}, especially when considering Pop~III stars which can be $\times 100$ more efficient in X-ray emissions than Pop~II stars \citep[][see discussion on Pop~III stars later]{2023MNRAS.521.4039S}. This would imply that heating is more efficient at higher-$z$, thus suppressing the 21-cm signal. \citet{2022MNRAS.513.5097K} quantify this effect by comparing Pop~II stellar models with two different $L_\text{X}/\text{SFR}-Z$ relations, one constant \citep{2013ApJ...776L..31F} and the other steeply increasing with metallicity \citep{2016MNRAS.457.4081B}. The suppression in the global signal in the latter model compared to the former is $\delta\sim\SI{25}{mK}$ for a signal of depth $\sim\SI{70}{mK}$. Thus, our lower limit at 95\% CI of a similar depth would decrease in such a model, but the \textit{prospect} for placing such a lower limit still remains. 
\end{enumerate}

\subsubsection{Radio emission at high-$z$}

Our model disfavours a strong radio background at high-$z$, inferring an upper limit of $T_\text{rad}(\nu_{21})\lesssim\SI{100}{K}$ at $z=6$, and $\lesssim\SI{50}{K}$ at $z>10$ (see Table~\ref{tab:TK_TS_constraints}), meaning that exotic astrophysical models of deep 21-cm signals are ruled out. This is conditional on our choice of radio SED (which is fixed to a synchrotron power law), and the same caveat that applies to the constant $L_X/\text{SFR}$ relation also applies to the $L_r/\text{SFR}$ relation. Galaxies at high-$z$ may have a steeper spectral index \citep[][favouring shallower signals]{2017A&A...602A...4D}, or higher radio luminosities (allowing for deeper signals). However, since radio emissions are largely unabsorbed in the IGM, it is hard to build up a large radio background without exceeding the constraints from LWA1/ARCADE2. As also pointed out in \citet{2025A&A...698A.152C}, the only way to allow for a large radio background at high-$z$ is to introduce an ad-hoc $z_\text{cutoff}$ below which radio emissions are quenched \citep[e.g.][]{2020MNRAS.499.5993R,2024ApJ...970L..25S} or to allow for them to be sourced by Pop~III galaxies that are sterilized by a growing Lyman-Werner background \citep[e.g.][]{2022MNRAS.511.3657M}, which leads onto our next caveat.

\subsubsection{Exclusion of Pop~III stars}
\label{sec:pop3_caveat}

There is growing interest in the modelling of Pop~III stars, and their observational prospects for JWST \citep[see, e.g.][and references within]{2024MNRAS.534..290L}. \simcode\ employs a flexible Pop~III model that allows for a range of Pop~III star formation efficiencies, IMFs \citep{2022MNRAS.516..841G,2025NatAs...9.1268G} and SED models \citep{2025MNRAS.541.3113L,2025arXiv250721764W}. As explained in Section~\ref{sec:astro_model}, the focus of our prior work \citep{2025MNRAS.542.2292D} was on Pop~II star formation constraints which this work builds on. Thus, we fix the Pop~III SFE to $f_{\star,\threei}=0.2\%$, the time for Pop~II star-formation to recover after Pop~III SNe to $t_\text{delay}=\SI{30}{Myr}$ \citep{2022MNRAS.514.4433M}, a log-flat IMF in the mass range $\SIrange{2}{180}{M_\odot}$, and the stellar modelling from \citet{2022MNRAS.516..841G}.

Including a range of Pop~III models in our analysis, as has been done in other recent works \citep[e.g.][]{2024MNRAS.529..519G,2024MNRAS.531.1113P,2025JCAP...10..047K}, may change the constraints. In particular, allowing for early WF coupling \citep[via high Pop~III SFE or strong Ly$\alpha$ emitting SED, both of which are poorly constrained; e.g.][]{2015MNRAS.448..654Y,2020MNRAS.495..123Q,2022MNRAS.511.3657M,2022MNRAS.516..841G}; early heating \citep[through an elevated $f_{\text{X},\threei}$, as opposed to a fixed $f_\text{X}$ for both populations; e.g.][]{2018MNRAS.478.5591M,2023MNRAS.521.4039S,2025NatAs...9.1268G,2025MNRAS.540..483V}; or a strong radio emitting population \citep[via $f_{\text{r},\threei}$; e.g.][]{2025A&A...698A.152C}, would indeed give rise to different 21-cm signals. Additionally, more extreme rotational Pop~III stars can even have a strong impact on the EoR by causing an early reionization scenario through their strongly ionizing SEDs \citep{2025MNRAS.541.3113L,2025arXiv250721764W}. This would suppress the signal by $\sim10-20\%$ at $z\sim15$.

We expect, conditional on the radio foreground fit, that SARAS~3 would disfavour efficient Pop~III star-formation since the global signal absorption trough would fall in the $z=15-25$ range, where the SARAS~3 non-detection lies \citep[e.g., see Figure 2 of][]{2024MNRAS.531.1113P}. Otherwise, like in our constraints, it would favour higher $V_c$ values pushing star-formation to sufficiently large atomic cooling halos that form at later times, thus making Pop~III star-formation at high-$z$ less likely. Recently, \citet{2024PhRvD.109d3523L} showed that including molecular cooling galaxies can decrease the inferred X-ray emission efficiency of atomic cooling galaxies when using HERA data. This allows for deeper, more extended absorption signals (within the SARAS~3 limits) which would bolster our optimistic predictions.

\subsubsection{Comment on reionization}
\label{sec:reionization}

There are a plethora of constraints on the reionization history of the Universe, some of which are plotted in Figure~\ref{fig:derived_constraints} (bottom panel). We have used the most conservative of these constraints in this work, a flat $3\sigma$ prior on the optical depth $\tau_\text{CMB}$ from \textit{Planck 2018} \citep{2020A&A...641A...6P}. The models presented here are thus agnostic to late or early reionization scenarios. We use a simple parameterization of the ionizing efficiency of galaxies $\zeta_\text{ion}$ that is fixed in both redshift and halo mass, despite the SFE being allowed to vary along both those variables. Whether such a model is realistic is outside the scope of this work, and we thus leave out the focus on ionization during EoR here. Nonetheless, since $\zeta_\text{ion} \propto f_\star f_\text{esc}$, the increased SFE at large halo masses may be compensated by a lower escape fraction \citep[as large halos are more resilient to SNe clearing channels of low density gas for ionizing photons to escape, e.g.][]{2016ApJ...833...84X,2020MNRAS.496.4342L,2024MNRAS.527.7924M}.

In future works, we hope to explore more flexible models for the ionizing properties of galaxies. This includes escape fractions (and their redshift/halo mass dependencies), clumping factors, ionizing emissivities --- all of which are subsumed into $\zeta_\text{ion}$ in this work due to their degenerate nature, but have been recently introduced in \simcode\ in \citep[][for Pop~III stars]{2025MNRAS.541.3113L}. This would allow us to not only explore varied reionization histories, but also build towards a more complete model of the astrophysical properties of galaxies at $z\gtrsim6$ by including various reionization datasets \citep[such as in][]{2021MNRAS.506.2390Q,2025PASA...42...49Q,2025JCAP...10..047K,2025arXiv250409725S}, giving tighter constraints on the vanishing of the global signal and power spectrum at the end of the EoR.

\section{Conclusions}
\label{sec:conclusions}

In this work, we constrain various properties of the neutral IGM at high-redshifts ($z \gtrsim 6$), including the kinetic temperature $T_\text{K}$, radio background temperature $T_\text{rad}(\nu_{21})$, the 21-cm spin temperature $T_\text{S}$, and consequently the 21-cm global signal and power spectrum. This follows from our previous work, \citet{2025MNRAS.542.2292D}, where we constrained the astrophysical properties of early Pop~II galaxies using a combination of multi-wavelength observations. This includes high-redshift UV luminosity functions (UVLFs) from HST and JWST, cosmic X-ray and radio backgrounds (CXB and CRB), 21-cm global signal non-detection from SARAS~3, 21-cm power spectrum upper limits from HERA, and CMB optical depth from \textit{Planck 2018} as a prior on our reionization history. Our main findings on the evolution of the IGM and the 21-cm signal are summarized below:
\begin{itemize}

    \item Using CXB data (in conjunction with UVLFs) and HERA upper limits, we constrain the evolution of the netural IGM temperature $T_\text{K}$ and find that the IGM is heated to at most ${T_\text{K}\sim \SI{7.7}{K}}$ at $z=15$ at 95\% credible interval (CI). At lower redshifts, it is heated to ${\SI{2.5}{K} \lesssim T_\text{K} \lesssim \SI{66}{K}}$ at $z=10$ and ${\SI{20}{K} \lesssim T_\text{K} \lesssim \SI{2078}{K}}$ at $z=6$ (as shown in Figure~\ref{fig:TK_TS_distributions} and Table~\ref{tab:TK_TS_constraints}).
    
    \item Using CRB data (in conjunction with UVLFs) and HERA upper limits, we constrain the radio background temperature at the rest-frame 21-cm wavelength, $T_\text{rad}(v_{21})$, to $\lesssim\SI{47}{K}$ at $z=15$ and ${\lesssim\SI{51}{K}}$ at $z=10$ (compared to ${T_\text{CMB}=\SI{43.6}{K}}$ and ${\SI{30.0}{K}}$ respectively), reaching a maximum of ${T_\text{rad}\lesssim\SI{101}{K}}$ at $z=6$ (compared to ${T_\text{CMB}=\SI{19.1}{K}}$). These constraints rule out exotic models of radio galaxies producing deep 21-cm signals at $z\gtrsim 15$.
    
    \item The combination of the above constraints in our joint analysis yields a constraint on the ratio of the 21-cm spin temperature to the radio background temperature, $T_\text{S}/T_\text{rad}\lesssim\SIrange{0.5}{0.8}{}$ at $z=12.5-15$. This has strong implications for the observability of the 21-cm signal.
    
    \item  We place the first lower limits on the 21-cm global signal absorption amplitude of $\SI{-201}{mK} \lesssim T_\text{21,min} \lesssim \SI{-68}{mK}$ at 95\% CI, at $z \approx 10-16$. Furthermore, we rule out a depth of $\gtrsim\SI{-33}{mK}$ at 99.99\% CI. The joint analysis also provides a lower limit on the 21-cm power spectrum at $z=15$ and $k=0.35\,h\text{Mpc}^{-1}$ of $\SI{8.7}{mK^2} \lesssim \Delta_{21}^2\lesssim \SI{197}{mK^2}$ at 95\% C.I, as shown in Figure~\ref{fig:21cm_constraints}.
    
    \item The above constraints assume that the same $L_X/\text{SFR}$ and $L_r/\text{SFR}$ relation holds for both Pop~II and Pop~III galaxies, and is fixed across redshifts. Furthermore, we do not include a flexible Pop~III stellar model in this work (instead fixing their parameters to a some fiducial values). Although we expect our constraints on the 21-cm signal to change with a broader model, the prospects for placing such lower bounds on the 21-cm signal using multi-wavelength synergies remain promising.
    
    \item The results are robust to the inclusion/exclusion of high-$z$ UVLF data from JWST, except for a marginal shift in the timing of the absorption trough to earlier epochs by $\Delta z\sim0.5$ if they are included. Similarly, the inclusion of SARAS~3 does not significantly change the constraints except allowing for marginally earlier absorption trough, by $\Delta z\sim0.5$, if excluded.
    
\end{itemize}

As the 21-cm community awaits independent verification of the EDGES global signal results \citep{2018Natur.555...67B}, and power spectrum experiments continue to place increasingly stringent upper limits \citep{2023ApJ...945..124H,2025A&A...698A.186M,2025ApJ...989...57N,2025MNRAS.542.2785M}, our work guides future observations by providing the first lower limit on the 21-cm signal, thus narrowing the discovery space. Our results also showcase the power of joint analyses using multi-wavelength datasets, which can be extended in the future to include Population~III parameters \citep[e.g.][]{2024MNRAS.531.1113P,2024PhRvD.109d3523L,2025A&A...698A.152C,2025NatAs...9.1268G,2025JCAP...10..047K}, or combined with other synergetic datasets from measurements of the neutral IGM fraction \citep[e.g.][]{2021MNRAS.506.2390Q,2025PASA...42...49Q,2025arXiv250409725S}, CMB observations \citep[e.g.][]{2018MNRAS.476.4025M,2020ApJ...899...40L,2023JCAP...08..017K,2025ApJ...991..195Z}, line intensity mapping experiments \citep[e.g.][]{2017ApJ...848...52H,2021MNRAS.507.2500M,2023ApJ...950...40S,2024ApJ...975..222F} or cosmic near-infrared background measurements \citep[e.g.][]{2014ApJ...790..148M,2021MNRAS.508.1954S,2025ApJ...981...92S}, to build a complete, self-consistent picture of the early Universe.

\section*{Acknowledgements}
JD would like to thank Adam Ormondroyd for his help in understanding and using \polychord, and the anonymous reviewer for their constructive feedback which has improved the manuscript.

JD acknowledges support from the Boustany Foundation and Cambridge Commonwealth Trust in the form of an Isaac Newton Studentship. TGJ and EdLA acknowledge the support of the Science and Technology Facilities Council (STFC) through grant
numbers ST/V506606/1 and a Rutherford Fellowship, respectively. HTJB would like to acknowledge support from the Kavli Institute for Cosmology Cambridge, and the Kavli Foundation. 

The simulations and analysis were performed under DiRAC project number APP15468 and APP54745 using the DiRAC@Durham facility managed by the Institute for Computational Cosmology on behalf of the STFC DiRAC HPC Facility (www.dirac.ac.uk). The equipment was funded by BEIS capital funding via STFC capital grants ST/P002293/1, ST/R002371/1 and ST/S002502/1, Durham University and STFC operations grant ST/R000832/1. DiRAC is part of the National e-Infrastructure.

\section*{Data Availability}

The 30,000 simulations performed using \simcode, the trained emulators, the \polychord\ chains, and other analysis scripts used in this work can all be made available upon reasonable request to the corresponding author.


\bibliographystyle{mnras}
\bibliography{21cm_constraints}



\appendix

\section{Special cases of the joint analysis}
\label{sec:appendix_joint_analysis}

\begin{figure*}
   \includegraphics[width=1\linewidth]{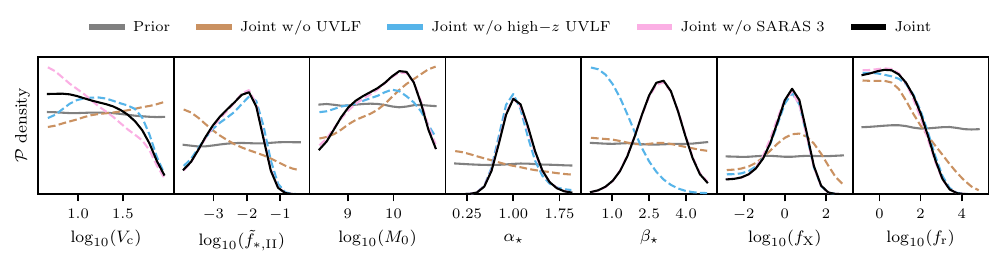}
    \caption{Same as Figure~\ref{fig:param_constraints}, but for special cases of the joint analysis where we exclude specific datasets. Qualitatively, the inclusion of UVLF data is crucial for constraining the SFE and for more accurate constraints on $f_\text{X}$ and $f_\text{r}$. The inclusion of high-$z$ UVLF data gives preference to a redshift-evolving SFE (i.e., $\beta_\star>0$). Finally, SARAS~3 data disfavours efficient star-formation in molecular cooling halos (i.e., low $V_c$).
    \label{fig:param_constraints_joint}}
\end{figure*}

\begin{figure*}
   \includegraphics[width=1\linewidth]{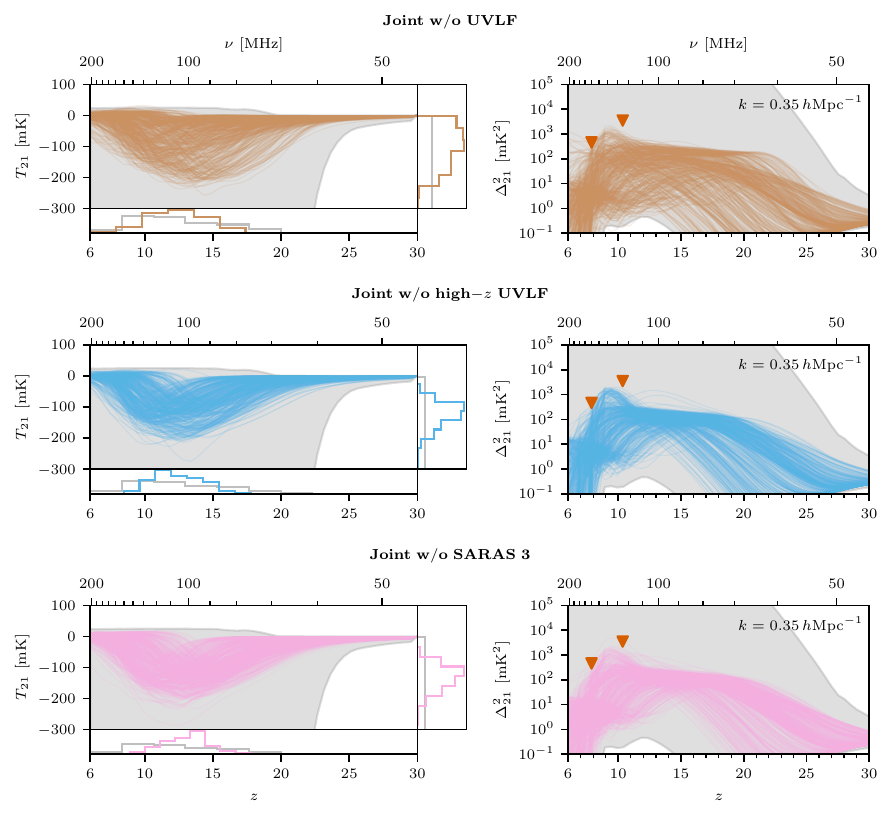}
    \caption{Same as the bottom row of Figure~\ref{fig:21cm_constraints}, but for special cases of the joint analysis where we exclude specific datasets.
    \label{fig:21cm_constraints_joint}}
\end{figure*}

Figure~\ref{fig:param_constraints_joint} shows the constraints on the astrophysical parameters from the joint analysis, similar to Figure~\ref{fig:param_constraints}, but for special cases where we exclude specific datasets, and Figure~\ref{fig:21cm_constraints_joint} shows the 21-cm global signal and power spectrum constraints.

\label{lastpage}
\end{document}